%% file: main_PRL_version.tex
\documentclass[aps,prl,twocolumn,nofootinbib]{revtex4-2}

\usepackage[dvipsnames]{xcolor}
\usepackage{hyperref}
\hypersetup{
  breaklinks=true,
  colorlinks=true,
  allcolors=BlueViolet,
}

\usepackage{graphicx} % inserting figures
\usepackage{svg}
\usepackage{amsmath}
\usepackage{physics,braket,bm}

\usepackage{xfrac}
\usepackage{enumerate}
\usepackage[normalem]{ulem} % for strikethrough

\usepackage{comment}

\newcommand \be{\begin{equation}}
\newcommand \ee{\end{equation}}

\newcommand \bea{\begin{eqnarray}}
\newcommand \eea{\end{eqnarray}}

\usepackage{amssymb}

\usepackage[inline]{enumitem}
\setlist[enumerate]{leftmargin=*}

\newcommand {\rsub}[1]{\textcolor{black}{#1}}

\newcommand{\UWM}{Department of Physics, University of Wisconsin-Madison,  Madison, WI, 53706}

\definecolor{mscolor}{rgb}{0,0.5,0.5}
\definecolor{mscolor}{rgb}{0,0.5,0.5}
\definecolor{cpcolor}{rgb}{0.4,0,0.8}
\definecolor{cpcolor}{rgb}{0.5,0,0.5}
\begin{document}

\title{Atom–photon Entanglement with a Single Trapped Cesium Atom}
\author{H. Hwang,
J. Moon,
F. Herzallah,
E. Oh,
A. Safari, and 
M. Saffman}
\affiliation{\UWM}
\date{\today}

\begin{abstract}
We demonstrate atom--photon entanglement using a single cesium atom trapped in an optical tweezer. 
Entanglement is generated by resonant excitation and subsequent spontaneous decay, which entangles the atomic Zeeman state with photon polarization. The emitted photon is coupled into a single-mode fiber, enabling measurement of the Bell-state fidelity. We obtain raw entanglement fidelity of 
${\mathcal F} = 0.942(16)$ and inferred fidelity of ${\mathcal F}_{\rm inf} = 0.962(26)$ after correcting independently characterized atom measurement errors. Compared with related free-space experiments using $^{87}$Rb, the Zeeman 
sub-structure of the relevant excited level in $^{133}$Cs requires the use of a single short excitation pulse in each entanglement attempt in order to suppress unwanted re-excitation. These results establish a free-space $^{133}$Cs atom--photon interface and provide a step toward dual-species Rb--Cs quantum networking.
\end{abstract}

\maketitle

%\tableofcontents

\noindent
\textcolor{blue}{\it Introduction} 
Quantum networks aim to interconnect spatially separated quantum processors and sensors using photonic links, thereby providing a route towards modular quantum information processing 
~\cite{Kimble2008,Wehner2018,Reiserer2015,Covey2023}.
In this architecture, each node combines long-lived stationary qubits for storage and processing with an interface to flying qubits that distribute entanglement across the network~\cite{Duan2001,Sangouard2011}. 
A fundamental resource in a quantum network is atom-photon entanglement which enables remote entanglement of stationary qubits, for example, through interference of  photons and Bell-state measurements~
\cite{
Hong1987,
Simon2003, 
Duan2003, 
Browne2003, 
Feng2003}. 
A central performance metric is therefore the rate and fidelity with which qubit--photon entanglement can be generated, collected, and delivered into a well-defined optical mode suitable for interference and Bell-state measurements
~\cite{
Volz2006,
Wilk2007b,
vanLeent2020,
Zhang2022,
Hartung2024,
LLi2025,
Safari2026}.
Remote entanglement has been realized on several platforms, including trapped ions
~\cite{Stephenson2020, Krutyanskiy2023, OReilly2025}, 
color centers~\cite{Pompili2021}, quantum dots~\cite{Stockill2017}, and neutral atoms~\cite{Welte2021, vanLeent2022}.
The generation rate and fidelity of remote entanglement links are largely determined by the light–matter interface at each node, by the coherence properties of stationary qubits, and by the stability and loss of the optical links.

Among these  platforms, neutral-atom tweezer arrays provide a scalable and reconfigurable approach with programmable geometries  enabling flexible connectivity  for large quantum registers~
\cite{Barredo2016,
Endres2016,
Muniz2025,
Bluvstein2026}.
Beyond static geometries, dynamic reconfiguration, atom transport, and reservoir-based refilling can enable sustained operation of large ordered arrays by replenishing lost atoms and maintaining a usable register for repeated communication attempts~
\cite{Pause2023,Norcia2024,
Gyger2024,Chiu2025}.
Heterogeneous dual-species neutral-atom systems offer additional capabilities for species-selective trapping, control, and measurement and differentiated roles with one species providing memory and processing, while the other serves as communication qubits \cite{Young2022}. This dual-species architecture enables mid-circuit readout with negligible optical crosstalk \cite{Singh2023, Miles2026} which is valuable for continuous operation of a network architecture  and will enable entanglement distillation protocols to be incorporated into network nodes \rsub{\cite{BLi2026}}.

Single neutral atom -- photon entanglement has so far been demonstrated with $^{87}$Rb in several free-space and cavity-based implementations
~\cite{Volz2006,Wilk2007b,vanLeent2020,Zhang2022,Hartung2024,Safari2026}
and with $^{171}$Yb atoms \cite{LLi2025}  in free space. While entanglement has been demonstrated between $^{133}$Cs ensembles containing many atoms and photons \cite{Chou2005}, we  demonstrate here for the first time a $^{133}$Cs atom--photon entanglement interface using a single atom in an optical tweezer. The atomic qubit is encoded in long lived hyperfine states and the optical qubit is encoded in the polarization basis.  Entanglement is demonstrated using the approach introduced in \cite{Volz2006} for $^{87}$Rb, but applied to the level structure of $^{133}$Cs. In addition we demonstrate mapping of the atomic qubit to internal states that extend the coherence time  by two orders of magnitude compared to the magnetically sensitive states that are natively prepared in the entanglement generation process. 
 We quantify the Bell-state fidelity and analyze the dominant error mechanisms, thereby establishing $^{133}$Cs as a viable atomic element  for future Rb--Cs dual-species quantum networks and modular quantum processors.

\begin{figure*}[!t]
\rsub{\includegraphics[width=\textwidth]{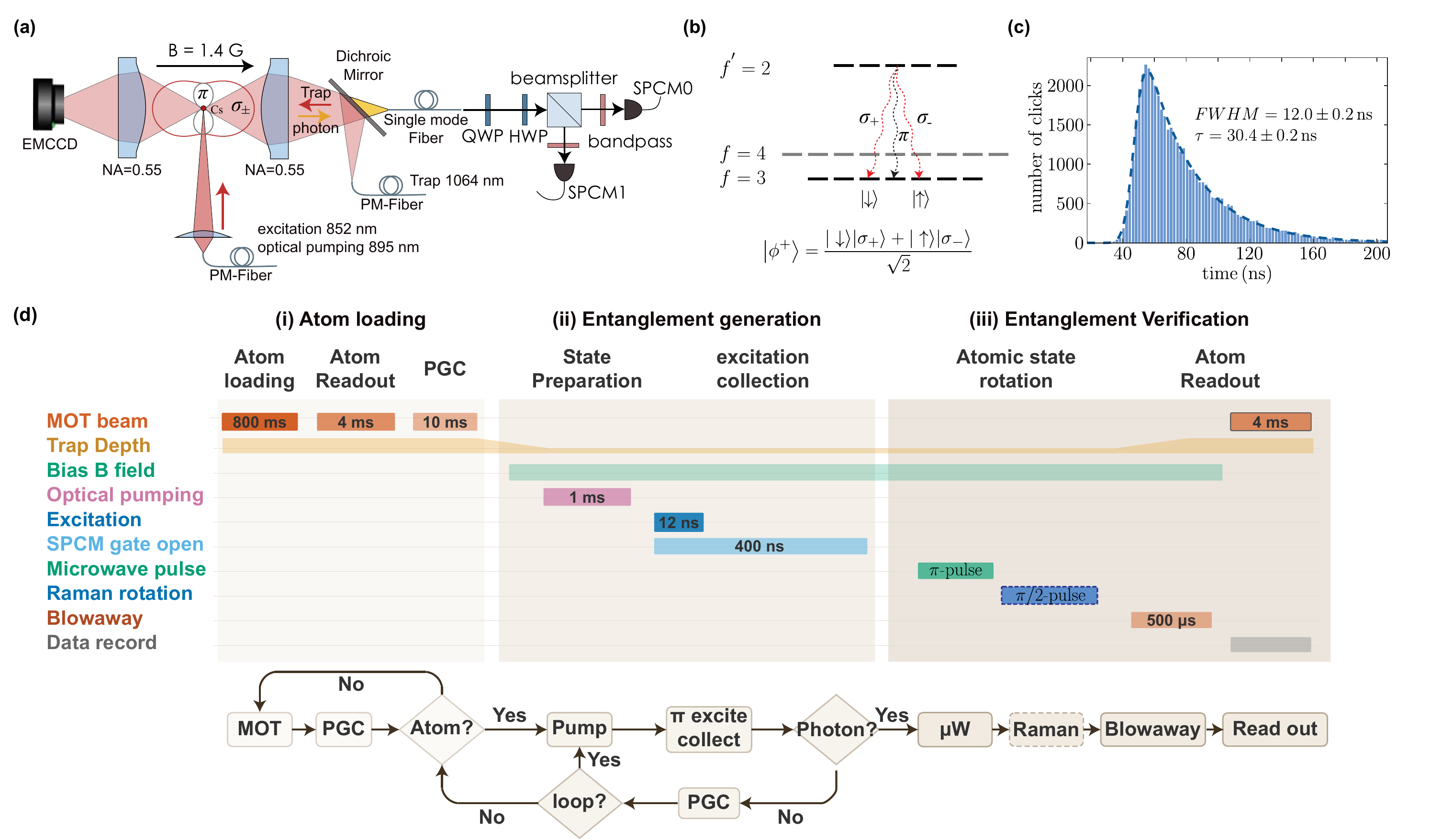}}
\caption{Optical setup and experimental sequence.
(a) Experimental setup used for $^{133}$Cs atom trapping, single-photon characterization, and atom--photon entanglement verification. 
\rsub{The two NA = 0.55 objective lenses have an effective focal length of 23.4 mm and are separated by approximately 37 mm between their facing surfaces. The magneto-optical trap (MOT) is formed between the objectives and spatially overlaps the optical tweezer. The single-lens symbols are simplified representations of the multi-element objective lenses.} 
For the $g^{(2)}$ measurement, a 50:50 non-polarizing beamsplitter is used, yielding $g^{(2)}(0)=0.096(66)$. For atom--photon entanglement, the beamsplitter is replaced by a polarizing beamsplitter (PBS). 
\rsub{(b) Cesium atom--photon decay channel leading to a heralded atom–photon Bell state. (c) Photon arrival-time histogram fitted by a Gaussian excitation pulse convolved with exponential decay, yielding a decay time of 30.4(2) ns.} 
\rsub{(d)} Experimental sequence for entanglement generation and verification. 
\rsub{In stage (i), a single atom is loaded using a MOT and polarization-gradient cooling (PGC). After confirming single-atom loading,}
in stage (ii) the atom is optically pumped to $\ket{f=3,m_f=0}$ and excited to $\ket{f'=2,m_f=0}$. Emitted photons with $\sigma_{\pm}$ polarization are  coupled into a single-mode fiber, and detected on an SPCM with total collection and detection efficiency $\eta=0.6\%$. Conditioned on a photon detection event, the atomic state is analyzed as described in the main text \rsub{.} If no photon is detected, the cooling–optical 
pumping–excitation sequence is repeated. After every three attempts of excitation, we check whether the atom is still trapped. If the atom remains, we continue until a photon is detected, otherwise a new atom is loaded.}
 \label{fig.figure1}
\end{figure*}

\textcolor{blue}{\it Single photon and entangled state generation} 
Our experiment uses a high-NA objective lens both to trap a single atom and to collect photons emitted from it in free space. As shown in Fig.~\ref{fig.figure1}(a), 1064 nm trapping light is  focused by an  objective lens (Jenoptik) with numerical aperture $\mathrm{NA}=0.55$ to form a single optical tweezer. Fluorescence from the trapped atom is collected by two identical objectives placed on opposite sides of the vacuum cell, one of which is also used to form the optical tweezer. Photons collected by one objective are coupled into a single mode optical fiber leading to single-photon counting modules (SPCM-AQRH-15-FC) for single-photon and atom--photon correlation measurements, whereas photons collected by the second objective are imaged onto an EMCCD camera to identify the spatial position of the trapped atom.  The trap beam waist ($1/e^2$ intensity radius) is 1.10(6)~$\mu$m, and the maximum photon collection efficiency of the optical system is 1.27\% (see \cite{Hwang2026SM} for details of the efficiency estimate).

Single photons are generated by preparing a $^{133}$Cs magneto-optical trap (MOT) and then loading a single atom into the optical trap. The atom 
%in 
\rsub{is}
pumped into the state $\ket{6s_{1/2},f=3,m_f=0}$
and is then excited to 
%$\ket{6p_{3/2},f=2,m_f=0}$
\rsub{$\ket{6p_{3/2},f'=2,m_f=0}$} 
with a fast optical pulse  as illustrated in Fig.~\ref{fig.figure1}(b) (see \cite{Hwang2026SM} for details).  Subsequent spontaneous decay  results in emission of $\sigma_+$, $\pi$, or $\sigma_-$ polarized photons together with  population of the atomic Zeeman states $\ket{6s_{1/2},f=3,m_f=-1,0,+1}$, respectively. The full atom--photon state therefore contains three components: 
$\ket{-1,\sigma_+}$,
$\ket{0,\pi}$ and 
$\ket{+1,\sigma_-}$.
However, the $\pi$-polarized component is strongly suppressed in the detected mode because of destructive interference of lobes of the dipole emission pattern when coupled into the single mode fiber ~\cite{Young2022}. As a result, a detected photon heralds the Bell state
\begin{equation}
\ket{\phi^+}=
\frac{|\downarrow\rangle|\sigma_+\rangle+|\uparrow\rangle|\sigma_-\rangle}{\sqrt{2}},
\label{sec2_bell}
\end{equation}
which is the atom--photon entangled state analyzed in this work. Here we have introduced the notation $\ket{\downarrow}=\ket{-1}$ and $\ket{\uparrow}=\ket{+1}.$

A key difference compared to similar experiments with  ${}^{87}\mathrm{Rb}$ \cite{Volz2006} that use excitation of $\ket{f'=0,m_{f'}=0}$ is that the excited level $\ket{6p_{3/2},f'=2}$ in ${}^{133}\mathrm{Cs}$ contains multiple Zeeman sub-states. Consequently, following optical pumping of ${}^{133}$Cs atoms, only a single short excitation pulse is applied in order to suppress re-excitation after spontaneous decay. Repeated excitation would populate unwanted Zeeman states and reduce the entanglement fidelity~\cite{Safari2026}.  Numerical modeling  \cite{Hwang2026SM} shows that a pulse duration of 30 ns would introduce an infidelity of approximately 3.5\% in the heralded Bell-state preparation due to re-excitation of $\ket{\uparrow}$ or $\ket{\downarrow}$ states. As shown in Fig.~\ref{fig.figure1}(\rsub{d}), we therefore use a single 12 ns excitation pulse for each entanglement attempt. After optical pumping and excitation, the SPCM gate is opened for 400~ns to detect an emitted photon. If a photon is detected, we proceed to the entanglement-verification sequence, including the atomic mapping pulse and state-selective measurement for entanglement verification. If no photon is detected, the atom is recooled and the optical-pumping and excitation sequence is repeated. After every three excitation attempts, we image the atom to check whether it remains trapped. If the atom is still present, the retry loop continues, but if the atom is lost the sequence returns to atom loading and recapture to continue the experiment until photon detection occurs.  

Collected photons are detected at the SPCMs at a rate of $0.3~\rm s^{-1}$.  The single-photon character of the source is verified by measuring
\begin{equation}
g^{(2)}(0)=\frac{P_{12}}{P_1P_2},
\end{equation}
where $P_1$ and $P_2$ are the probabilities of a count on the two detectors after a 50:50 non-polarizing beamsplitter, and $P_{12}$ is the coincidence probability~\cite{HanburyBrown1956}. We obtain $g^{(2)}(0)=0.096(66)$, consistent with near single photon emission.
In addition, we record time-tagged detection events with a resolution of 1 ns using the ARTIQ-based experimental control system.  As shown in Fig.~\ref{fig.figure1}(\rsub{c}), the photon arrival-time histogram is well fit by the convolution of a Gaussian excitation pulse with an exponential spontaneous-decay response. From this fit we extract a decay time of 30.4(2) ns, in agreement with the known lifetime of the $6p_{3/2}$ state. The same system is used to generate and synchronize the experimental pulse sequence shown in Fig.~\ref{fig.figure1}(\rsub{d}).

\begin{figure*}
\centering
\includegraphics[width=.8 \textwidth]{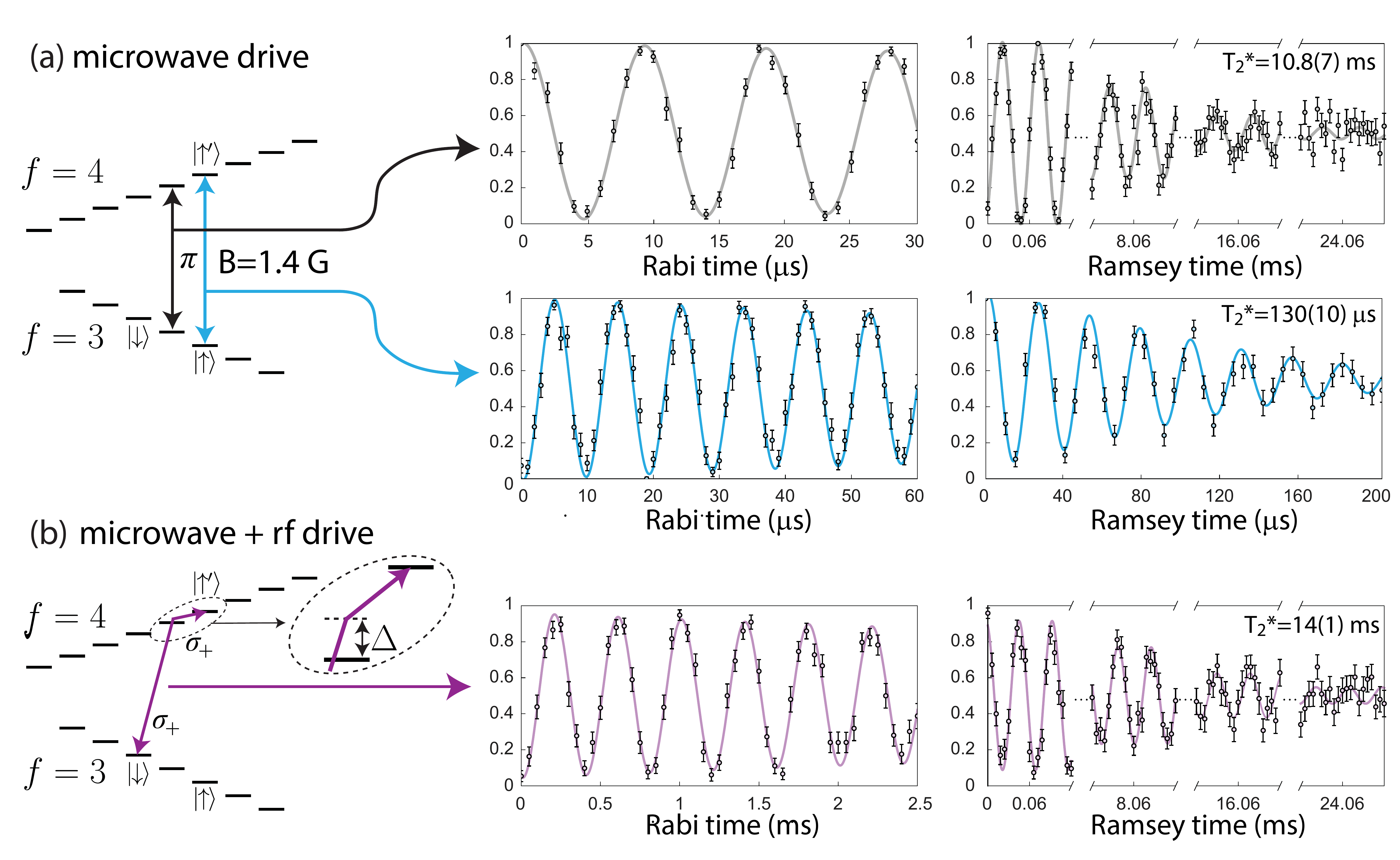}
\caption{Coherent control and coherence of the ground-state hyperfine manifold of a trapped $^{133}$Cs atom. 
(a) Left: level diagram illustrating $\pi$-polarized microwave control at a bias field $B=1.4~\mathrm{G}$ for the transitions between clock states (black) and $\ket{\uparrow}\!\leftrightarrow\!\ket{\uparrow'}$ (blue). Center: measured microwave Rabi oscillations as a function of pulse length. Measurement shows Rabi frequency of $\Omega_{\text{clock}}=2\pi\times 109.0(9)~\mathrm{kHz}$ (black) and $\Omega_{\ket{\uparrow}\!\leftrightarrow\!\ket{\uparrow'}}=2\pi\times 104.04(59)~\mathrm{kHz}$ (blue)  Right: Ramsey fringes versus free-evolution time. Fits to a decaying sinusoid yield $T_2^*=10.8(7)~\mathrm{ms}$ (black) and $T_2^*=130(10)~\mathrm{\mu s}$ (blue). 
(b) Two-photon stimulated Raman control between $\ket{\downarrow}$ and $\ket{\uparrow'}$ with $T_2^*=14(1)~\mathrm{ms}$ using $\sigma^+$--$\sigma^+$ microwave and rf fields (purple). A 9.2 GHz microwave field and a 1 MHz rf field generated by a horn and antenna are blue detuned by $\Delta=2\pi\times ~125$ kHz from
$|6s_{1/2},f=4,m_f=0\rangle$. The  effective two-photon Rabi frequency
$\Omega_{\mathrm{eff}}=\Omega_{\mu \text{W}}\Omega_{\text{rf}}/2\Delta=2\pi\times 2.497(6)$ kHz. The microwave Rabi frequency was measured to be
$\Omega_{\mu \text{W}} = 2\pi\times 13.5(2)$ kHz, and the rf Rabi frequency was estimated as $\Omega_{\mathrm{rf}}\simeq 2\pi\times 46.2$ kHz using the expression for $\Omega_{\rm eff}$.  }
 \label{fig.figure2}
\end{figure*}

\noindent
\textcolor{blue}{\it Atom-photon entanglement verification} 
 The emitted photon and the single atom ideally form the Bell state $\ket{\phi^+}$ of 
 Eq. (\ref{sec2_bell}). 
 Because this Zeeman state qubit is sensitive to
magnetic field fluctuations, the atom--photon state acquires a shot-to-shot
random relative phase, which reduces the observable coherence.
The magnetic dephasing is reflected in the Ramsey data shown in Fig.~\ref{fig.figure2}.
The magnetically less sensitive clock state superposition between $\ket{f=3, m_f=0}$ and $\ket{f=4,m_f=0}$ exhibits a coherence time of
$T_2^*=10.8(7)~\rm  ms$, primarily limited by atomic temperature and the differential light shift between hyperfine levels due to the trap light.  The  magnetically sensitive superposition of $|\uparrow\rangle=\ket{f=3,m_f=1}$ and $|\uparrow'\rangle=\ket{f=4,m_f=1}$  shows a much shorter coherence
time of $T_2^*=130(10)~\mu\rm s$ dominated by magnetic noise. Although we do not directly measure $T_2^*$ for the $\ket{\uparrow}-\ket{\downarrow}$ qubit we expect it has a similarly short coherence time. 

To reduce this source of error we transfer the atomic state $\ket{\uparrow}$ to
$\ket{\uparrow'}$ using a 5~$\mu$s duration 
mapping pulse implemented with a 9.2 GHz microwave field.  
\rsub{The resulting mapped-qubit transition is first-order insensitive to magnetic-field fluctuations at $B \simeq 1.4~\mathrm{G}$. The measured field fluctuations are approximately $2.5~\mathrm{mG}$ \cite{Hwang2026SM} and therefore make a negligible contribution to the atomic-qubit phase noise. After mapping, as shown in Fig.~\ref{fig.figure2}(b), the qubit has an atomic coherence time of $T_2^* = 14(1)~\mathrm{ms}$.}
We note that this is more than $4\times$ longer than the coherence observed in the corresponding $\ket{\downarrow}-\ket{\uparrow'}$ basis with Rb atoms \cite{Safari2026}. The improved coherence compared to that work is due to use of further detuned trap light which reduces differential light shift effects and the larger hyperfine splitting of ${}^{133}$Cs which reduces sensitivity to magnetic noise. 

A complete characterization of the prepared Bell state $\ket{\phi^+}$ requires tomography of the $4\times4$ atom--photon density matrix. Instead, we efficiently determine the Bell-state fidelity from the diagonal Pauli-basis correlations $XX$, $YY$, and $ZZ$, using the expression \cite{Bennett1996}
\begin{equation}
\label{eq:fidelity_xxyyzz}
\mathcal{F}=\frac{1+\langle X_{\rm A} X_{\rm P}\rangle-\langle Y_{\rm A} Y_{\rm P}\rangle+\langle Z_{\rm A} Z_{\rm P}\rangle}{4},
\end{equation}
where A and P denote the atomic and photonic qubits, respectively.

\begin{figure}[t!]
\includegraphics[width=\columnwidth]{ 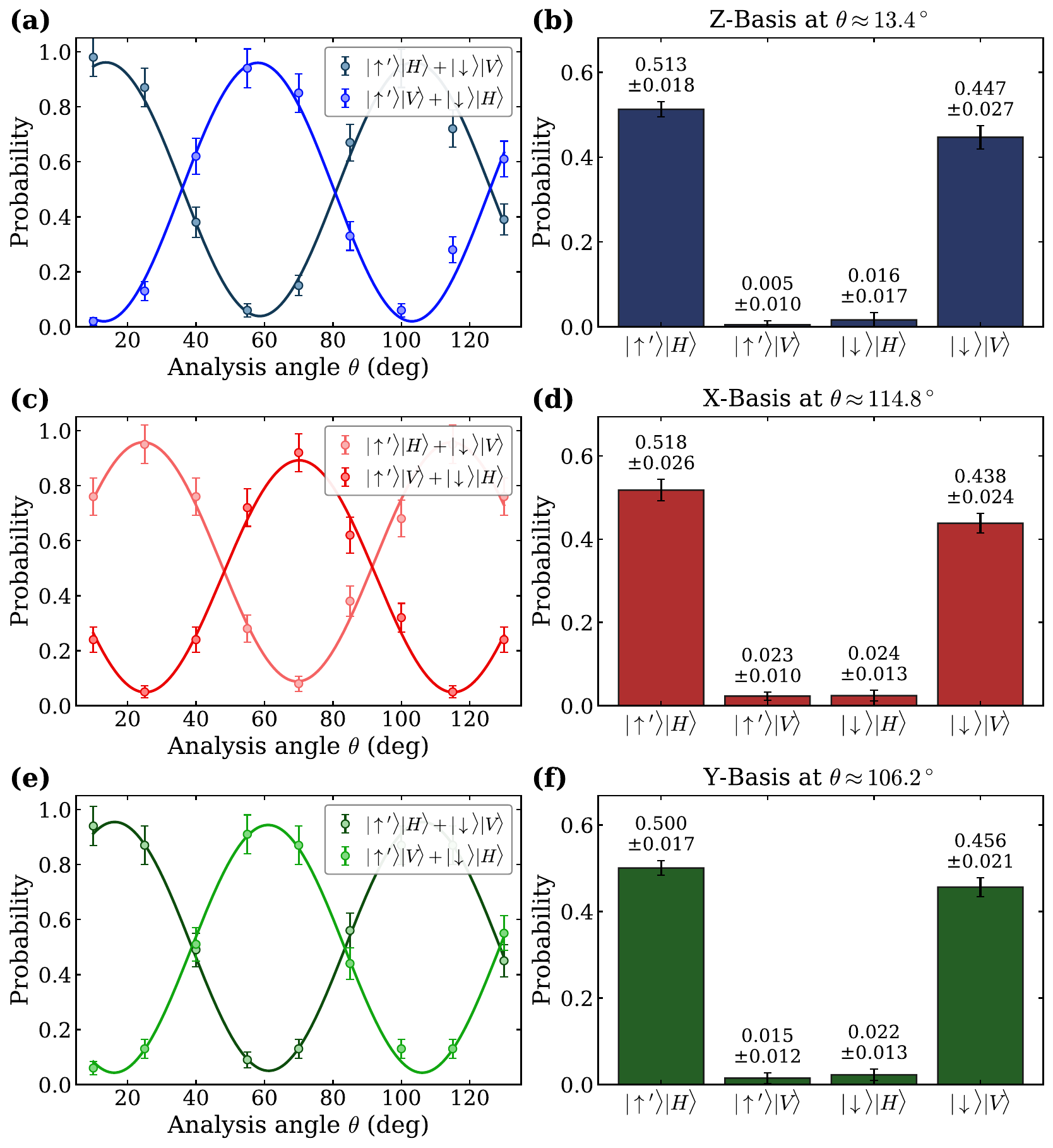}
\caption{(a), (c), and (e) show the atom--photon parity oscillations measured in the $Z$, $X$, and $Y$ bases, respectively, as a function of the photon HWP angle $\theta$. Each data point is the average of 100 measurements, with error bars indicating $\pm$ one standard deviation. The even-parity outcomes correspond to $\ket{\uparrow'}\ket{H}+\ket{\downarrow}\ket{V}$, while the odd-parity outcomes correspond to $\ket{\uparrow'}\ket{V}+\ket{\downarrow}\ket{H}$. Solid curves are independent sinusoidal fits to the measured probabilities. (b), (d), and (f) show the reconstructed joint probabilities evaluated at the HWP angles that maximize the parity-oscillation \rsub{contrast} in the $Z$, $X$, and $Y$ bases, respectively.}
 \label{fig.figure3}
\end{figure}

We perform Bell state tomography by fixing the measurement basis of the atom and subsequently aligning the photon analysis basis. For measurements in the $Z$ basis the atomic state is measured by mapping $\ket{\uparrow}$ to $\ket{\uparrow'}$ followed by  blow-away of the $\ket{\uparrow'}$ state using a single beam resonant with  the $\ket{6s_{1/2},f=4}\rightarrow\ket{6p_{3/2},f'=5}$ transition and then an atom occupancy measurement.    For the photonic qubit, the emitted photon is coupled into a single-mode fiber, followed by a quarter-wave plate (QWP), a half-wave plate (HWP), and a polarizing beam splitter (PBS) before detection by one of two single-photon counting modules (SPCMs)
%(see Fig.~\ref{fig.figure1}a )
\rsub{as shown in Fig.~\ref{fig.figure1}(a)}. 
Because the fiber applies an unknown  unitary transformation to the photon polarization, the absolute photonic measurement basis is not known \textit{a priori}. However, under the hypothesis that the system is highly entangled, setting the atomic basis dictates the expected correlations. We therefore establish the corresponding photonic basis empirically by scanning the QWP and HWP angles to maximize the contrast of the atom--photon parity oscillations. Figure~\ref{fig.figure3}(a) shows the measured parity oscillation as a function of the HWP angle $\theta$ in the $Z$ basis.
\rsub{Using a temperature stabilized enclosure, the optimal photon-analyzer waveplate angles remained stable to within $0.5^{\circ}$ over a day, indicating that low-frequency thermal drift in the fiber-induced polarization transformation is negligible in the present setup. For long fiber links, such low-frequency noise can induce larger polarization drifts, which can be compensated using fixed-polarization reference light and active polarization feedback \cite{Kucera2024}.}

Unlike the $Z$-basis measurement, the $X$- and $Y$-basis measurements require a coherent rotation between $\ket{\downarrow}$ and $\ket{\uparrow'}$. Because these states differ by $\Delta m_f=2$, this rotation is implemented using a $\pi/2$ pulse on a two-photon microwave - radio frequency transition, as shown in Fig.~\ref{fig.figure2}. The $Y$-basis measurement is performed by shifting the phase of the two-photon drive by $\pi/2$ relative to the $X$-basis measurement. Any fixed phase offsets in the atomic analysis basis---such as the differential AC Stark shift induced by the microwave and RF fields---are naturally absorbed when we scan the photon waveplates to maximize the parity-oscillation contrast. Figures~\ref{fig.figure3}(c) and (e) show the resulting parity oscillations in the $X$ and $Y$ bases.
\rsub{Alternatively, the phase offsets could be calibrated independently by applying the same mapping and analysis sequence to a separately prepared coherent atomic superposition and determining them via Ramsey interferometry.}

%\rchange{To evaluate the atom--photon entanglement fidelity, we extract the joint expectation values from}{To determine the optimal measurement basis, we perform}
\rsub{To determine the optimal measurement basis, we perform}
independent sinusoidal fits to the parity oscillations in Fig.~\ref{fig.figure3}(a), (c), and (e) \rsub{using the functional form $P_j(\theta) = C_j [1 + V_j \cos(4\theta - \phi_j)]$, where $j \in \{{\rm even}, {\rm odd}\}$}.  
The even and odd parity states are  
$\ket{\rm even}=\ket{\uparrow'}\ket{H}+\ket{\downarrow}\ket{V}$ 
and $\ket{\rm odd}=\ket{\uparrow'}\ket{V}+\ket{\downarrow}\ket{H}$, respectively. 
Fitting the even and odd parity curves independently accounts for baseline asymmetries, ensuring the 
\rsub{contrast}
is not artificially constrained to be constant across the scan. 
\rsub{To evaluate the atom--photon entanglement fidelity, we apply the same unconstrained functional form to perform four additional independent fits on the individual joint measurement outcomes. The expectation values in Eq.~(\ref{eq:fidelity_xxyyzz}) are obtained from the four reconstructed joint probabilities shown in Fig.~\ref{fig.figure3}(b), (d), and (f), evaluated at the corresponding angles of maximum parity contrast.} The QWP and PBS convert the circularly polarized photons to the $H,V$ basis states which are directed to the photon detectors. 

An experimental detail arises in the $Y$-basis measurement. The maximum contrast was observed at a HWP angle $\theta$  that projects the photon onto $-Y$, so we measure the $Y_{\rm A}(-Y_{\rm P})$ correlations. 
Consequently, the extracted contrast directly represents $-\langle Y_{\rm A} Y_{\rm P} \rangle$, which substitutes into the fidelity expression (\ref{eq:fidelity_xxyyzz}) without further sign correction. 
\rsub{From these four independent joint probability fits, we reconstruct the expectation values} 
$\langle Z_{\rm A} Z_{\rm P}\rangle=0.939(34)$, $\langle X_{\rm A} X_{\rm P}\rangle=0.909(39)$, and $-\langle Y_{\rm A} Y_{\rm P}\rangle=0.919(32)$ at the HWP angles listed in Fig.~\ref{fig.figure3}. Using these values in Eq.~(\ref{eq:fidelity_xxyyzz}), we determine the atom--photon entanglement fidelity to be
$ \mathcal{F}=0.942(16). $ 

We also determined a lower bound on the entanglement fidelity using the
diagonal elements of the density matrices $\rho$ and $\tilde{\rho}$ measured
 in the $Z$ and $Y$ bases, respectively~\cite{Blinov04}
\begin{comment}
\begin{equation}
\label{eq.Fidelity_bound}
\begin{split}
{\mathcal F} \ge \frac{1}{2}\Big(
&\rho_{\downarrow H,\downarrow H}
+\rho_{\uparrow' V,\uparrow' V}
-2\sqrt{\rho_{\downarrow V,\downarrow V}\rho_{\uparrow' H,\uparrow' H}}
\\
&+\tilde{\rho}_{\downarrow H,\downarrow H}
+\tilde{\rho}_{\uparrow' V,\uparrow' V}
-\tilde{\rho}_{\downarrow V,\downarrow V}
-\tilde{\rho}_{\uparrow' H,\uparrow' H}
\Big).
\end{split}
\end{equation}
\end{comment}
\rsub{
\begin{equation}
\label{eq.Fidelity_bound}
\begin{split}
{\mathcal F} \ge \frac{1}{2}\Big(
&\rho_{\uparrow' H,\uparrow' H}
+\rho_{\downarrow V,\downarrow V}
-2\sqrt{\rho_{\uparrow' V,\uparrow' V}\rho_{\downarrow H,\downarrow H}}
\\
&+\tilde{\rho}_{\uparrow' H,\uparrow' H}
+\tilde{\rho}_{\downarrow V,\downarrow V}
-\tilde{\rho}_{\uparrow' V,\uparrow' V}
-\tilde{\rho}_{\downarrow H,\downarrow H}
\Big).
\end{split}
\end{equation}}
Using the atom--photon correlated populations
$P(\uparrow',H)=\rho_{\uparrow'H,\uparrow'H}$,
$P(\downarrow,H)=\rho_{\downarrow H,\downarrow H}$,
$P(\uparrow',V)=\rho_{\uparrow'V,\uparrow'V}$, and
$P(\downarrow,V)=\rho_{\downarrow V,\downarrow V}$ taken from the $Z$- and $Y$-basis data shown in Fig.~\ref{fig.figure3}(b) and (f), we evaluate Eq.~(\ref{eq.Fidelity_bound}) and obtain a lower bound on the atom--photon entanglement fidelity,
\begin{equation}
     {\mathcal F} \ge {\mathcal F}_{\mathrm{low}} = 0.931(25).
\end{equation}

The dominant error sources limiting the entanglement fidelity are dephasing during the mapping pulse to the magnetically insensitive state for the  $X$- and $Y$-basis parity measurements ($\epsilon_{\mathrm{coh}}=0.0305(15)$) and errors in the atomic state measurement ($\epsilon_{\mathrm{state}}=0.02(2)$). These errors and additional smaller effects are detailed in \cite{Hwang2026SM}. Following the logic used in Ref.~\cite{Safari2026}, we quote an inferred fidelity obtained by correcting only for the independently calibrated atomic state detection error, which gives ${\mathcal F}_{\mathrm{inf}}\approx0.962(26)$.

\noindent
\textcolor{blue}{\it Summary and outlook}
We have realized atom--photon entanglement with a single trapped $^{133}$Cs atom and 
\rsub{measured}
an entanglement fidelity of $\mathcal{F}=0.942(16)$.
Compared to analogous  experiments with $^{87}$Rb, atom--photon entanglement in $^{133}$Cs is complicated by the larger Zeeman-state manifold of the relevant excited state. This multilevel structure introduces additional decay channels and intrinsic errors associated with unwanted re-excitation. Despite these constraints, our results show that atom--photon entanglement with comparable fidelity can be generated  with $^{133}$Cs atoms. Furthermore we demonstrate more than $4\times$ improvement of the qubit coherence relative to related recent work in $^{87}$Rb \cite{Safari2026}.
\rsub{Excitation-pulse shaping, faster mapping to the magnetically
insensitive basis states, and Raman sideband cooling could further improve
the Bell-state fidelity. In parallel, reducing atom-loading overhead
\cite{Pause2023,Gyger2024,Chiu2025}, together with cavity-enhanced
photon collection \cite{Hartung2024}, could in principle raise the
detected entanglement-generation rate beyond $100~\mathrm{s}^{-1}$.}

This demonstration establishes $^{133}$Cs as a feasible atomic element for atom--photon interfacing for quantum networking. More broadly, it provides a key ingredient for Rb-Cs dual-species neutral-atom architectures \cite{Young2022,Anand2024,Miles2026}, where different atomic species can be used for communication and memory/processing functions with reduced optical crosstalk and enhanced operational flexibility. This dual-species architecture with in-place mid-circuit measurements is particularly well suited for future entanglement distillation experiments. Our work thus represents a step towards heterogeneous Rb--Cs quantum network nodes and scalable neutral-atom quantum networking.

 The authors thank Seonghyun Park and Akshat Prakash for technical assistance with the experimental apparatus. This material is based upon work supported by NSF Award No. 2016136 for the QLCI center Hybrid Quantum Architectures and Networks, the U.S. Department of Energy Office of Science National Quantum Information Science Research Centers as part of the Q-NEXT center, and NSF Award No. 2228725.

\bibliography{qc_refs,rydberg,saffman_refs,atomic,optics,supplemental_refs}

%\end{document}

% \pagebreak 
% .
% \newpage

\onecolumngrid

\include{SM_body.tex}

\end{document}

%% file: SM_body.tex
\noindent
{\Large Supplemental material for:}

\begin{center}

{\Large\bf Atom–photon Entanglement with a Single Trapped Cesium Atom}

H. Hwang,
J. Moon,
F. Herzallah,
E. Oh,
A. Safari, and 
M. Saffman

Department of Physics, University of Wisconsin-Madison,  Madison, WI, 53706

\end{center}

\date{\today}

\setcounter{page}{1}
\setcounter{section}{0}
\setcounter{figure}{0}
\setcounter{equation}{0}
\setcounter{table}{0}

\renewcommand{\thepage}{SM.\arabic{page}}
\renewcommand{\theequation}{SM.\arabic{equation}}
\renewcommand{\thesection}{SM.\arabic{section}}
\renewcommand{\thefigure}{SM.\arabic{figure}}
\renewcommand{\thetable}{SM.\arabic{table}}

\section{Atom-photon entanglement experiments}

Table \ref{tab.atom_photon_entangle} lists demonstrations of entanglement between single neutral atoms and photons. The fidelity achieved here with $^{133}$Cs atoms is higher than all but one of the previous demonstrations with other species. The success probability is comparable to other free space experiments that used lenses for photon collection, but lower than the efficiency of the parabolic mirror based demonstration with $^{87}$Rb \cite{Safari2026}.

\begin{table*}[!b]
\footnotesize
\caption{Demonstrations of neutral atom-photon entanglement. Success probability is the probability of detecting a photon for each attempted excitation of the atom. Reported fidelity values are measured results without correction for separately characterized error sources.}
\label{tab.atom_photon_entangle}
\centering

\resizebox{\textwidth}{!}{%
\begin{tabular}{|l|l|l|c|c|c|}
\hline
Year-group & Description & photon encoding & matter qubit & 
success probability & fidelity \\
\hline
2006-Weinfurter \cite{Volz2006} & lens, $\rm{NA}=0.38$, $\rm{NA}_{eff}=0.29$ & polarization & $^{87}$Rb & $0.0005$ & 0.87\\
2007-Rempe \cite{Wilk2007b} & cavity, $C=1.28$ & polarization & $^{87}$Rb & 0.013 & 0.86\\
2020-Weinfurter \cite{vanLeent2020} & lens, $\rm{NA}=0.5$ & polarization & $^{87}$Rb & $0.0075$ & $0.897^c$ \\
2022-Weinfurter \cite{Zhang2022} & lens, $\rm{NA}=0.5$ & polarization & $^{87}$Rb & Trap 1(2)$^b$: $0.00598(0.00144)$ & $0.952(0.941)$ \\
2024-Rempe \cite{Hartung2024} & cavity, $C=1.66$ & polarization & $^{87}$Rb & 0.33 & 0.86\\
2025-Covey \cite{LLi2025} & lens, $\rm{NA}=0.63^a$ & time bin & $^{171}$Yb & 0.003 & $0.90$ \\
2026-Saffman \cite{Safari2026} & parabolic mirror, $\rm{NA}=0.61$ & polarization & $^{87}$Rb & $0.029$ & $0.93$ \\
2026-(this work) & lens, $\rm{NA}=0.55$ & polarization & $^{133}$Cs & 0.006 & $0.942(16)$ \\
\hline
\end{tabular}%
}

\vspace{2pt}
\parbox{\textwidth}{\footnotesize
a) J. Covey, private communication.
b) Success probability for Trap 2 is lower due to 50\% loss from 700 m fiber.
c) This fidelity is for $5~\mathrm{m}$ fiber length without quantum frequency conversion.
}
\end{table*}

\section{State preparation and photon generation}
 
 Single photons are generated by preparing a $^{133}$Cs magneto-optical trap (MOT) and then loading a single atom into the optical trap. 
 \rsub{The atom is} 
 prepared in the state $\ket{6s_{1/2},f=3,m_f=0}$ by optical pumping using $\pi$-polarized 895 nm light driving the $\ket{6s_{1/2},f=3}\rightarrow\ket{6p_{1/2},f'=3}$ transition together with all polarizations of 
852 nm light on the $\ket{6s_{1/2},f=4}\rightarrow\ket{6p_{3/2},f'=4}$ transition provided by the MOT beams.
Before optical pumping, the trap depth is adiabatically reduced from 0.96~mK to 0.24~mK to minimize the vector light shift from the tightly focused tweezer and  position-dependent differential AC Stark shifts~\cite{Thompson2013b}. We achieve an optical pumping fidelity of 0.991(7) with a pumping time of 1 ms.

After state preparation, the atom is excited by a resonant $\pi$ pulse on the $\ket{6s_{1/2},f=3,m_f=0}\rightarrow\ket{6p_{3/2},f'=2,m_f=0}$ transition, as illustrated in Fig.~1(b) in the main text. Subsequent spontaneous decay  results in emission of $\sigma_+$, $\pi$, or $\sigma_-$ polarized photons together with  population of the atomic Zeeman states $\ket{6s_{1/2},f=3,m_f=-1,0,+1}$, respectively. The full atom--photon state therefore contains three components: 
$\ket{-1,\sigma_+}$,
$\ket{0,\pi}$ and 
$\ket{+1,\sigma_-}$.
However, the $\pi$-polarized component is strongly suppressed in the detected mode because of  destructive interference of lobes of the dipole emission pattern when coupled into the single mode fiber ~\cite{Young2022}. As a result, a detected photon heralds the Bell state
\begin{equation}
\ket{\phi^+}=
\frac{|\downarrow\rangle|\sigma_+\rangle+|\uparrow\rangle|\sigma_-\rangle}{\sqrt{2}},
\rsub{\label{sec2_bell_supply}}
\end{equation}
which is the atom--photon entangled state analyzed in this work. Here we have introduced the notation $\ket{\downarrow}=\ket{-1}$ and $\ket{\uparrow}=\ket{+1}.$

\begin{figure}[!t]
\includegraphics[width=.6 \columnwidth]{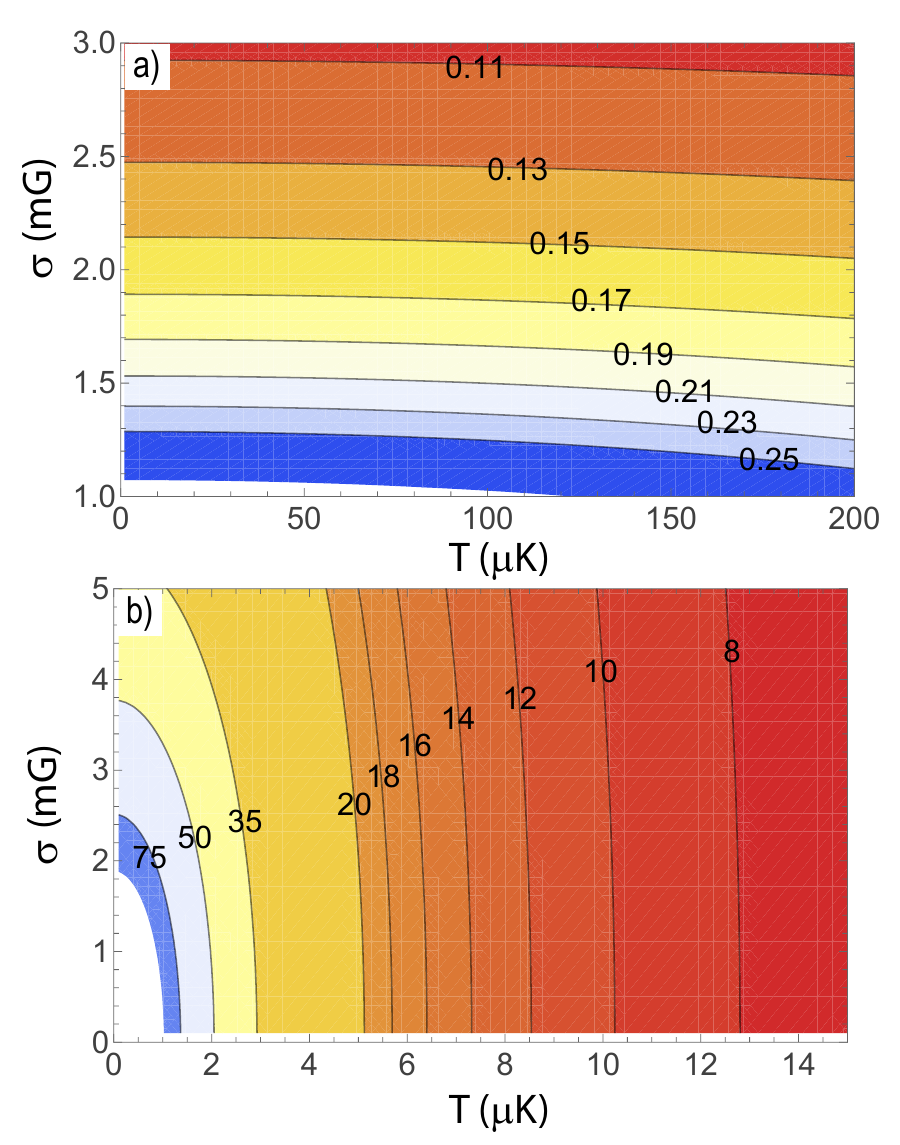}
 \caption{\rsub{Calculated Ramsey coherence time as a function of atom temperature and magnetic field noise at a bias field of $B_0=1.4~\rm G$ and a trap wavelength of 1064 nm.  We assume  Gaussian magnetic field noise with standard deviation of $\sigma$. Contours are labeled with the coherence time in ms. a) Calculated coherence of the $\ket{3,1}\rightarrow\ket{4,1}$ transition which is primarily sensitive to magnetic noise. The observed coherence time of $T_2^*=0.13~\rm  ms$ indicates $\sigma=2.47~\rm mG$. b) Calculated coherence of the $\ket{3,0}\rightarrow\ket{4,0}$ clock transition. The observed coherence time of $T_2^*=10.8 ~\rm ms$ indicates a temperature of $T=9.5~\mu\rm K$.}}
 \label{fig.SM_fig_AtomT}
\end{figure}

\section{Atom temperature measurement}
\label{app.atomTemp}

\rsub{The temperature of the trapped atoms is calculated by comparing the measured  coherence time of the clock transition from the Ramsey data shown in Fig.~\ref{fig.figure2}(a) with the  calculated  Ramsey coherence as a function of both atomic temperature and magnetic field noise. The results are shown as a two-dimensional color map in Fig.~\ref{fig.SM_fig_AtomT}. Following Ref.~\cite{Kuhr2005}, the calculation is based on thermal averaging of the differential light shift experienced by the atom in the trap, together with the residual sensitivity of the clock transition to magnetic-field fluctuations \cite{Saffman2011}. For each temperature, the atomic motion is treated within the thermal distribution of the trapping potential, which gives a distribution of light shifts and therefore a corresponding inhomogeneous dephasing of the Ramsey signal. Magnetic field noise is included as an additional Gaussian broadening of the transition frequency. The coherence of the mapping transition (see Fig. \ref{fig.figure2}(a)) primarily depends on the magnetic field 
which implies noise of $\sigma=2.47~\rm mG$. This agrees reasonably well with the estimate of 2 mG measured with a magnetometer. }

\rsub{Note that the noise of 2.47 mG about the $B_0=1.4~\rm G$ magic point implies a  correction to the qubit frequency after mapping to the field insensitive states of only 2.6 mHz. This value is found from the quadratic dependence of the transition frequency on the magnetic field near the magic value. This additional dephasing is negligible compared to the observed  $T_2^*=14~\rm  ms$ coherence time which is limited by the atomic temperature.  }

\rsub{
Since the clock qubit at our bias field is only weakly sensitive to magnetic field fluctuations, the calculated clock transition coherence time depends predominantly on the atomic temperature over the relevant parameter range. By comparing the measured value $T_2^* = 10.8~\mathrm{ms}$ to the calculated map, we infer an atomic temperature of approximately $9.5~\mu\mathrm{K}$.}

\section{Entanglement error budget}

Information concerning estimation of the error sources 
\rsub{listed}
in Table \ref{tab.errors} is provided in this section. 
Unless otherwise noted, the quoted values are conservative estimates intended to identify the dominant contributions to the measured atom--photon infidelity.

\paragraph*{Atom dephasing}

A dominant contribution to the infidelity arises from atomic dephasing during the $X$- and $Y$-basis parity measurements. Because the ARTIQ-based control sequence requires a finite time interval for frequency updates and for scheduling the mapping and $\pi/2$ analysis pulses, the experimental sequence includes an approximately $7~\mu$s delay between photon detection and the mapping pulse, followed by a further $\sim125~\mu$s during the $X$- and $Y$-basis analysis sequence. 

The atom-dephasing contribution was estimated from the measured Ramsey coherence times of the two atomic qubits involved in the $X$-basis analysis. 
From Fig. 2(a) of the main text,
\rsub{we estimated} the magnetic-field-sensitive qubit $(|\downarrow\rangle,|\uparrow\rangle)$ has a fitted coherence time of 
\rsub{$\tau_{\mathrm{sens}}\ge 130(10)~\mu\mathrm{s}$}, while from 
\rsub{Fig.~\ref{fig.figure2}(b)} of the main text, the mapped qubit $(|\downarrow\rangle,|\uparrow'\rangle)$ has a fitted coherence time of $\tau_{\mathrm{map}}=14(1)~\mathrm{ms}$.

Using the experimental timing sequence, we take a pre-mapping delay of $t_{\mathrm{pre}}=7~\mu\mathrm{s}$ in the magnetic-sensitive basis and a post-mapping interval of $t_{\mathrm{post}}=25~\mu\mathrm{s}+100~\mu\mathrm{s}=125~\mu\mathrm{s}$ in the mapped basis. 
From the Ramsey data, the remaining coherence is estimated as
\[
C=\exp\!\left(-\frac{t_{\mathrm{pre}}}{\tau_{\mathrm{sens}}}\right)
  \exp\!\left(-\frac{t_{\mathrm{post}}}{\tau_{\mathrm{map}}}\right).
\]
Substituting the measured values gives $C=0.939(3)$, which corresponds to an atom--photon fidelity reduction of
\begin{equation}
\epsilon_{\phi}=\frac{1-C}{2}=0.0305(15).
\end{equation}
The factor of $1/2$ is  introduced since the Z basis  parity is not affected by the  dephasing error, only the X and Y parities, Propagating the fit uncertainties of the two coherence times gives $\epsilon_{\phi}=0.0305\pm0.0015$. 
We therefore assign an atom-dephasing contribution of $0.0305\pm0.0015$ to the error budget. 
This estimate is conservative, since the full $230~\mu\mathrm{s}$ analysis pulse is included in the dephasing interval.

\paragraph*{Atomic state measurement}
The atomic-state-measurement contribution was obtained experimentally from the state-selective detection sequence used in the main experiment. 
State detection is based on a blow-away pulse that removes atoms in the $|6s_{1/2},f=4\rangle$ level, followed by fluorescence readout of the remaining atom.  The infidelity of the state-selective measurement was calibrated experimentally by preparing atoms in $|6s_{1/2},f=3\rangle$ and $|6s_{1/2},f=4\rangle$ and comparing the outcomes of the blow-away/readout sequence.
This  yielded an estimated contribution of $0.02\pm0.02$.
This error affects all of the $X,Y,$ and $Z$ basis parity measurements.

\paragraph*{Leakage during excitation}
The third contribution is due to leakage errors from excitation of $m_f\ne 0$ Zeeman states in 
\rsub{$6p_{3/2},f'=2$.} 
This error is estimated to be 0.017 on the basis of simulations presented in the next section.

\begin{table}[!t]
\caption{Main sources of atom-photon entanglement infidelity with their corresponding estimated contributions.}
\label{tab.errors}
\centering
\begin{tabular}{|l| c| }
\hline
source of error &  contribution  \\
\hline
atom dephasing &  $0.0305 \pm \rsub{0.0015}$ \\
atomic state measurement &  $0.02 \pm 0.02$ \\
leakage during excitation &  $\sim0.017$ \\
imperfect optical pumping & $0.009\pm0.007$ \\
photon detection noise & \rsub{$<6\times10^{-3}$} \\
atom basis preparation &  $<3\times 10^{-3}$ \\
waveplate rotation error &  $<3\times10^{-3}$ \\
excitation polarization &  $<1\times10^{-4}$ \\
\hline
quadrature sum  $\left(\sum_j \epsilon_j^2\right)^{1/2}$ & 0.042 \\
\hline
\end{tabular}\\
\end{table}

\paragraph*{Atomic basis preparation.}
The atomic-basis-preparation contribution accounts for imperfect preparation of the analysis basis, primarily due to uncertainty in the resonance condition of the mapping pulse to the magnetic-field-insensitive state. 
Using the fitted resonance-frequency uncertainty relative to the Rabi frequency of the nominal $\pi$ mapping pulse, the corresponding transfer error is estimated to be below $3\times10^{-3}$. 
We therefore conservatively assign $\epsilon_{\mathrm{prep}}<3\times10^{-3}$.

\paragraph*{Photon detection noise.}
The photon-detection-noise contribution was estimated from the ratio of the background count probability to the true herald probability. 
In the present experiment, the 
\rsub{photon detection success probability} is approximately $0.6\%$, while the background count probability in the photon collection window of 400 ns is at the level of $2\times10^{-5}$ per attempt based on 
\rsub{the dark counts of each SPCM.} 
Treating background-triggered events as uncorrelated false heralds gives an atom--photon fidelity reduction at the $10^{-3}$ level. 
To remain conservative, we assign an upper bound of 
\rsub{$\epsilon_{\mathrm{bg}}<6\times10^{-3}$}
\rsub{as the sum of the contributions from the two SPCMs, each estimated from the gate-open time $\times$ dark-count rate divided by the total photon-detection success probability.}
\rsub{Replacing the SPCMs with SNSPDs would reduce the dark-count rate from $50~\mathrm{s}^{-1}$ to $3~\mathrm{s}^{-1}$ at 852~nm, thereby reducing the infidelity by a factor of approximately $50/3\simeq17$, from  $0.6\%$ to $0.04\%$ for the two detector channels used in the experiment.}

\paragraph*{Excitation polarization.}
Imperfect excitation polarization can populate unwanted Zeeman transitions and reduce the atom--photon entanglement fidelity. 
We estimate this contribution from two effects: the finite extinction ratio of the excitation optics and the position-dependent tilt of the local quantization axis caused by the tightly focused optical tweezer. The excitation beam is nominally $\pi$ polarized with an extinction ratio of approximately $10^{4}:1$, corresponding to a direct wrong-polarization intensity fraction of about $10^{-4}$. 
A second contribution arises from the spatially varying ellipticity of a tightly focused optical tweezer, which produces an effective vector light shift and a position-dependent local quantization axis. 
\rsub{Following Ref.~\cite{Thompson2013b}} 
and using our trap parameters (NA$=0.55$, $\lambda=1064$ nm, trap depth $0.24$ mK, atom temperature $\sim10~\mu$K), we estimate an rms local-axis tilt of order $1^\circ$. 
This corresponds to a wrong-polarization admixture of order a few $\times10^{-4}$. 
Combining this with the direct optics impurity gives a total wrong-polarization fraction of order a few $\times10^{-4}$, and we therefore conservatively bound the corresponding atom--photon infidelity by $\epsilon_{\mathrm{pol}}<1\times10^{-4}$.

\paragraph*{Imperfect optical pumping.}
The optical pumping fidelity is approximately $0.991(7)$ in the present experiment calculated by taking the ratio of the pumping and de-pumping time constants. Residual population outside the target state contributes to the final atom--photon error budget with $\epsilon_{\mathrm{pump}}=0.009\pm0.007$.

Additional smaller contributions arise from atomic basis preparation, photon detection noise, waveplate rotation error, and excitation polarization. These contributions were estimated from the uncertainty in the microwave resonance, the precision of the waveplate settings, and the position-dependent fictitious magnetic field generated by the tightly focused optical tweezer together with the atomic temperature~\cite{Thompson2013b}.

Taken together, these independently estimated contributions are consistent with the measured atom--photon entanglement fidelity, 
${\mathcal F}=0.942(16)$ and the lower bound of ${\mathcal F}_{\mathrm{low}}=0.931(25)$. 
Following the logic used in Ref.~\cite{Safari2026}, we quote an inferred fidelity in the main text obtained by correcting only for the independently calibrated atomic state detection error, which gives ${\mathcal F}_{\mathrm{inf}}\approx0.962(26)$. The other contributions listed in Table~\ref{tab.errors}, such as atom dephasing and imperfect basis preparation, are treated as physical sources of infidelity and are therefore not corrected in the inferred value.

\section{Errors due to excitation of other Zeeman states} 
\label{Supple_sec1}

 \rsub{The $\ket{6p_{3/2},f'=2}$} upper level in the excitation scheme used for entanglement generation has five Zeeman sublevels, while we only wish to excite the $m_f=0$ sublevel. The additional  sublevels can be populated  due to a double-excitation process. Here, we analyze the errors caused by occupation of unwanted states during the excitation and relaxation process. 
\rsub{A difference from the commonly used $^{87}\mathrm{Rb}$ scheme arises from the excited-state Zeeman structure. In $^{87}\mathrm{Rb}$, excitation to the $f'=0,m_f'=0$ state is followed by spontaneous decay to the $m_f=\pm1$ ground states. These states cannot be re-excited by the original $\pi$-polarized field because the $f'=0$ manifold does not contain $m_f'=\pm1$ sublevels. Re-excitation following photon emission is forbidden, in principle. In contrast, after the $^{133}\mathrm{Cs}$ atom decays to $m_f=\pm1$, the remaining $\pi$-polarized excitation pulse can drive the atom to the corresponding $f'=2,m_f'=\pm1$ states. A subsequent excitation and decay can leave the final atomic state uncorrelated with the first emitted photon, thereby reducing the atom--photon entanglement fidelity. This difference motivates the numerical optimization of the excitation-pulse duration presented below.}

First, we solve the Lindblad Master equation to trace the population of the density matrix element in each state. There are a total of 28 sublevels involved in the calculation (7 states in $|f=3\rangle$, 5 states in $ |f'=2\rangle$, 7 states in $ |f'=3\rangle$, and 9 states in $ |f'=4\rangle$). An additional  sink state is added to account for decay to $|f=4\rangle$. With the $29\times29$  density matrix, we solve the Lindblad Master equation

\begin{equation}
        \frac{d\rho}{dt} = -\frac{i}{\hbar}[H(t), \rho] + \sum_k (L_k\rho L_k^{\dagger} - \frac{1}{2}\{L_k^{\dagger}L_k, \rho\})
        \label{eq.lindblad}
\end{equation}
The interaction term in the Hamiltonian $H(t)=H_0 + H_{\rm int}$ is modeled by the elements of the off-diagonal matrix mapped by a Gaussian excitation pulse, where $\langle f', m_{f'}|H_{\rm int}|f,m_f\rangle=\frac{\hbar\Omega(t)}{2}\times C^{fm_f}_{f'm_{f'}1\Delta m}$ with $\Omega(t)=\Omega_p e^{-\frac{(t-t_0)^2}{2\sigma^2}}$, and $C^{fm_f}_{f'm_{f'}1\Delta m}$  the corresponding Clebsch-Gordan coefficient for $\Delta m=m_f-m_{f'}.$ The peak Rabi frequency for the stretched transition $\ket{f=I+1/2,m_f=f}\rightarrow \ket{f'=I+3/2, m_{f'}=f'}$ is $\Omega_p$ and $I$ is the nuclear spin. 
The Rabi frequency $\Omega_p$ and width $\sigma$ are chosen to achieve a $\pi$ pulse area on resonance. 
\rsub{The Gaussian envelope was chosen to reproduce the experimentally measured excitation pulse, whose temporal profile recorded with a photodetector was approximately Gaussian.} 
Jump operators ${L_k}$ in the equation are defined as 
\begin{equation}
    L_k = \sqrt{\Gamma_{|f,m_{f}\rangle\leftarrow\ket{f',m_{f'}}}}  \ket{f,m_f}\bra{ f',m_{f'}}
\end{equation}
where the decay rate is
\begin{equation}
\Gamma_{|f,m_{f}\rangle\leftarrow|f',m_{f'}\rangle} = \Gamma\, b_{f\leftarrow f'}\left(C_{f'm_{f'}1q}^{fm_f}\right)^2   
\end{equation}
with $\Gamma$ the total decay rate of the excited state,  $b_{f\leftarrow f'}=(2j'+1)(2f+1)\left|\begin{Bmatrix} j & I & f \\ f' & 1 & j' \end{Bmatrix}\right|^2$  the branching ratio between hyperfine manifolds $f'\rightarrow f$, and $q=-1,0,1$ the spherical  component of the emitted photon.  

\begin{figure}[!t]
\centering
\includegraphics[width=1.0\textwidth]{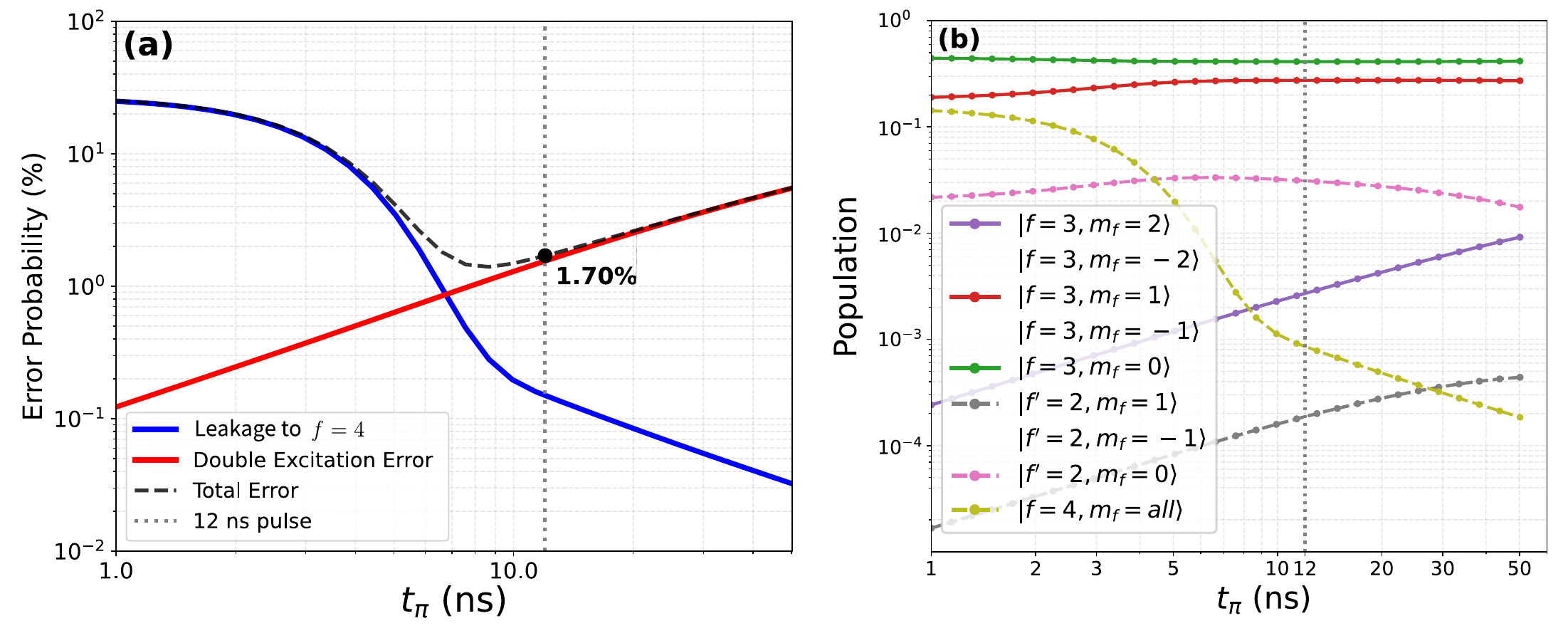}
\caption{Errors during atomic excitation. (a) Entanglement state generation infidelity from two dominant mechanisms. For short pulses, there is leakage to $\ket{6s_{1/2},f=4}$ shown by the blue curve.   For long pulses there are double excitation errors shown by the red curve.   The total simulated error, shown by the black dashed curve, has an optimum near $t_{\pi}\simeq 8~\mathrm{ns}$. In the experiment, we use $t_{\pi}=12~\mathrm{ns}$ due to limited laser power, indicated by the vertical dotted line, for which the simulated state-generation infidelity is approximately $1.7\%$.
(b) Simulated final atomic-state populations after the excitation pulse as a function of $t_{\pi}$. Several curves overlap because the corresponding symmetric Zeeman sublevels have nearly identical populations.}
\label{fig.appendix1}
\end{figure}

The simulation accounts for two different sources of infidelity. The first is that a short $\pi$ pulse implies a large peak Rabi frequency which causes undesired coupling to the levels with $f'=3,4.$ These levels can decay back to $f=4$ causing preparation of an erroneous atomic state.  The second type of error occurs when the pulse is long enough such that the excited atom can decay and be excited again before the end of the pulse, which may lead to preparation of an erroneous atomic state or an allowed state ($\ket{\downarrow}$ or $\ket{\uparrow}$) that is not correlated with the photon in the desired Bell superposition.
Both effects are accounted for by numerically solving Eq. (\ref{eq.lindblad}). 
\rsub{For the fixed-area Gaussian pulses considered here, the peak Rabi frequency scales as $\Omega_p\propto1/t_\pi$. At short pulse durations, the large peak Rabi frequency and broad spectral bandwidth produce strong off-resonant excitation of neighboring hyperfine manifolds. Because the initial $\pi$-polarized transition $\ket{f=3,m_f=0}\rightarrow \ket{f'=3,m_f'=0}$ is forbidden, the dominant direct leakage state is $\ket{f'=4,m_f'=0}$, resulting in the sharp increase in the leakage contribution as $t_\pi$ decreases. 
In contrast, as the pulse duration becomes comparable to the approximately $30~\mathrm{ns}$ lifetime of the cesium $6p_{3/2}$ state, the probability that the atom spontaneously decays while the excitation field is still present increases. For the fixed-area pulses considered here, the average excited-state population during the pulse remains approximately similar, so the probability of spontaneous decay before the end of the pulse increases approximately in proportion to $\Gamma t_\pi$. After such a decay, the remaining portion of the pulse can re-excite the atom. Consequently, the re-excitation contribution increases approximately linearly with pulse duration.}

In Fig. \ref{fig.appendix1} we show the error probability as a function of the $\pi$ pulse duration $t_\pi$, keeping the pulse area fixed for each value of $t_\pi$. That is, a shorter pulse requires higher peak laser power, while a longer pulse requires lower peak power. One can clearly see the dependence of the two error mechanisms on the  pulse duration. For short pulses, the error is dominated by power-broadening-induced leakage to the $f=4$ ground level, because the larger spectral bandwidth increases off-resonant excitation of $f'\ne 2$. On the other hand, for longer pulses, the double-excitation probability increases, and it becomes the dominant error source. The simulation predicts an optimal pulse duration of approximately 8 ns. In the experiment, we used $t_\pi=12$ ns, for which the simulated infidelity is $ \sim1.7$ \%.
\rsub{For pulse durations far beyond the range shown in Fig.~\ref{fig.appendix1}, the fixed pulse area results in a smaller peak Rabi frequency, reducing the spectral overlap with the Zeeman-detuned $\ket{f=3,m_f=\pm1}\rightarrow \ket{f'=2,m_f'=\pm1}$ transitions, so the remaining pulse becomes less effective at re-exciting the atom and the calculated re-excitation contribution eventually decreases. However, such long pulses would require a longer SPCM detection window, increasing the probability of dark-count-induced false-herald events. Therefore, such long pulse durations do not correspond to a useful operating regime for atom--photon entanglement generation.}
\rsub{Further improvement may be possible through optimized pulse shaping, for example using a Blackman--Harris envelope or a waveform obtained through GRAPE optimization, which could suppress off-resonant excitation from the $f=3$ ground-state to unwanted excited hyperfine states with $f'=4$.}

\section{Photon collection efficiency} \label{Supple_sec2}
To simulate the overall single-photon coupling efficiency of the experiment, we need to model the dipole emission pattern, the photon-collection efficiency, the fiber-coupling efficiency, and the thermal distribution of the atom \cite{garthoff2021efficient,Young2022}.
First, starting from the dipole emission and collection efficiency, the electric field from an oscillating dipole is given by $\mathbf{E}(\mathbf{r}) \propto (\mathbf{e}_r \times \mathbf{e}_p )\times \mathbf{e}_r$, where $\mathbf{e}_p$ is the direction of the dipole orientation and $\mathbf{e}_r$ is the unit vector along $\bf r$. Assuming the dipole is located at the focal point of the objective lens, the collection efficiency into an objective with a given NA is obtained by integrating the dipole radiation pattern over the solid angle subtended by the objective's aperture:
\begin{align}
    \eta^{(p)}_{\rm col}(\mathbf{r'}=0)=\frac{3}{8\pi}\int_{\Omega_{NA}} d\Omega \; r^2|\mathbf{E}^{(p)}(\mathbf{r})|^2
\end{align}

where $p$ is the polarization of the emitted electric field. The resulting photon collection efficiencies, depending on the polarization, are
\begin{align*}
    &\eta^{\sigma_\pm}_{\rm col} = \frac{1}{2} - \frac{3}{8}\cos\left({\theta_{\rm max}}\right)-\frac{1}{8}\cos^3\left({\theta_{\rm max}}\right)\\
    &\eta^{\pi}_{\rm col} = \frac{1}{2} - \frac{3}{4}\cos\left({\theta_{\rm max}}\right)+\frac{1}{4}\cos^3\left({\theta_{\rm max}}\right)
\end{align*}

where $\theta_{\rm max}=\sin^{-1}(\text{NA})$ is the half-angle of the acceptance cone of the lens.

\rsub{The dipole field of the emitted photon is mapped onto the principal plane of the objective, which is approximated as a thin lens, before evaluating its overlap with the Gaussian fiber mode. The thin lens applies a phase profile induced by path length difference of $\Delta(x,y)=\sqrt{x^2+y^2+f^2}-f$, where $f$ is the effective focal length of the objective, which transforms the spherical wavefront into a collimated wavefront. The dipole field at the objective is limited by the objective aperture. Consequently, the photon field after the objective becomes}
\begin{align*}
    \mathbf{E}^{(p)}_{\rm photon}(\mathbf{r}-\mathbf{r'}) &= \mathbf{E}^{(p)}(\mathbf{r}-\mathbf{r'})\cdot e^{-ik\Delta(x,y)}\cdot\Theta(A_{ap}) \\
    \Theta(A_{ap})&=\begin{cases} 
1, & \text{for } x, y \in A_{\text{ap}} \\ 
0, & \text{for } x, y \notin A_{\text{ap}} 
\end{cases}
\end{align*}
where $A_{ap}$ is the area of the aperture.
The single-mode fiber mode is approximated by a $\text{TEM}_{00}$ Gaussian mode, given by
\begin{align*}
    \mathbf{E}^{(q)}_{\rm fiber}(\mathbf{r},w_{\rm col})=\frac{1}{\sqrt\frac{\pi}{2}w_{\rm col}}e^{-\frac{x^2+y^2}{w_{\rm col}^2}}\begin{pmatrix}{} q_x \\ q_y \\0\end{pmatrix}
\end{align*}
\rsub{where $w_{\rm{col}}$ is the waist of the collimated Gaussian fiber mode.}
The coupling efficiency is then given by the normalized mode-overlap integral
\small
\begin{align}
    \eta^{(p,q)}_{cp}(\mathbf{r'}, w_{\rm col}) = \frac{|\int_{A_{\rm ap}}dA\,\mathbf{E}^{(p)}_{\rm photon}(\mathbf{r}-\mathbf{r'})\cdot\mathbf{E^*}^{(q)}_{\rm fiber}(\mathbf{r},w_{col})|^2}{N^{(p)}_{\rm photon}(\mathbf{r'})N^{(q)}_{\rm fiber}(w_{col})}
\end{align}
\normalsize
where $N^{(p)}_{\rm photon}(\mathbf{r'})$, $N^{(q)}_{\rm fiber}(w_{col})$ are normalization factors. 
\par Lastly, we must account for the thermal distribution of the atomic positions. Assuming that the trapped atom occupies a thermal distribution within a harmonic potential, its position distribution becomes
\begin{align}
    P_{\rm th}(r,z,T)=\frac{1}{(2\pi)^{3/2}\sigma_z\sigma_r^2}e^{-\frac{r^2}{2\sigma_r^2}}e^{-\frac{z^2}{2\sigma_z^2}}
\end{align}
with $\sigma_i=\sqrt{\frac{k_BT}{mw_i^2}}$ where $w_r$ and $w_z$ denote the transverse and longitudinal trap frequencies, respectively.

\begin{figure}[!t]
\centering
\includegraphics[width=1\columnwidth]{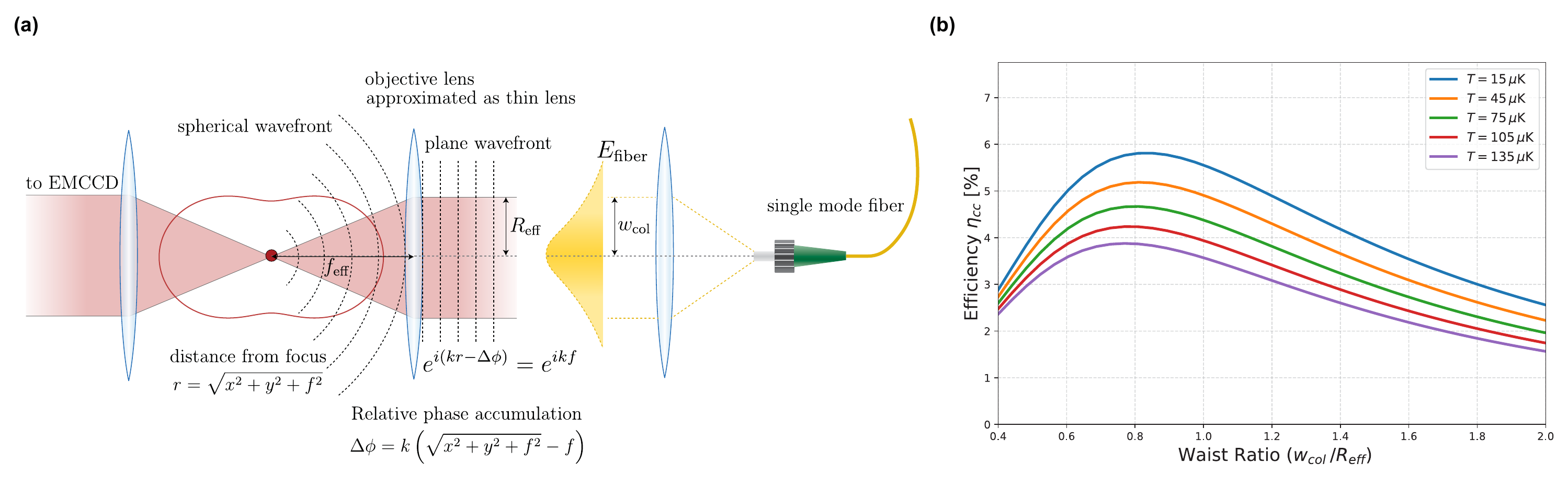}
\caption{\rsub{(a) Schematic of the photon collection and fiber-coupling model. The collection objective is approximated as a thin lens with effective focal length $f_{\rm eff}$ and effective aperture radius $R_{\rm eff}$. The lens removes the position-dependent phase of the spherical wavefront, producing a plane wavefront that is overlapped with the Gaussian fiber mode of waist $w_{\rm col}$. (b)} Simulation of the collection- and coupling efficiency of the dipole emission from the trapped $^{133}$Cs atom for several atomic temperatures. Calculations for lens $NA=0.55$, and trap depth $U_{\rm trap}=k_B \times 200 ~\mu\rm K$. $R_{\rm eff}$ is the effective aperture radius  defined by 
\rsub{$R_{\rm eff}=f_{\rm{eff}}\cdot \tan{(\theta_{\rm max}})$}}
\label{fig.appendix2}
\end{figure}

Now we combine the equations above to calculate the thermally averaged combined collection and coupling efficiency, defined as
\begin{equation}
    \eta^{(p,q)}_{\rm cc}(w_{col}) = b_p\int d^3r'\;P_{\rm th}(\mathbf{r'},T)\cdot\eta^{(p)}_{\rm col}(\mathbf{r'})\cdot\eta^{(p,q)}_{cp}(\mathbf{r'},w_{\rm col})
\end{equation}
where $b_p$ is the branching probability for emitting polarization $p$, with $\sum_p b_p = 1$($b_{\sigma^+}=b_{\sigma^-}=\frac{2}{7}$ and $b_{\pi}=\frac{3}{7}$). In the simulation, we only assume the collection of $\sigma^{\pm}$ polarization channel.
By substituting the numerical values for our experiment ($NA=0.55$, 
\rsub{$w_{\rm col}=6.96$ mm,} 
$\lambda_{\rm trap}=1064\, \text{nm}$, $w_{\rm trap, input}= 12\,\text{mm}$, $U_{\rm trap}=200\;\mu$K, $T_{\rm atom}=5\;\mu$K), we obtain a final collection and coupling efficiency of 
$\eta_{\mathrm{cc}}^{(\sigma)}=4.135\%$.

\rsub{Using the same simulation model, with the emitted field replaced by the
$\pi$-dipole field, we obtained
$\eta_{\mathrm{cc}}^{(\pi)}=0.004\%$. This quantitative value shows that destructive interference of the
$\pi$-polarized photon in the single-mode fiber results in a suppression factor
$S=\eta_{\mathrm{col}}'^{(\pi)}/\eta_{\mathrm{cc}}^{(\pi)}
\simeq 200$, where $\eta_{\mathrm{col}}'^{(\pi)}$ is the objective collection
efficiency of the $\pi$-polarized photons including the branching ratio of the
atomic decay channel, and a $\sigma$- to $\pi$-photon coupling ratio
$R=\eta_{\mathrm{cc}}^{(\sigma)}/\eta_{\mathrm{cc}}^{(\pi)}
\simeq 10^3$. However for a misaligned imaging system, the contributions from the opposite lobes of the $\pi$-dipole emission pattern do not fully cancel in the Gaussian fiber mode because a small transverse displacement $\delta$ makes their contributions unequal.
The resulting residual field amplitude is approximately proportional to $\delta$, so the coupled $\pi$-photon probability increases approximately as $\delta^2$. We obtain
$(S,R)\simeq(26,100)$ and $(11,30)$ for
$\delta=0.25~\mu\mathrm{m}$ and $0.50~\mu\mathrm{m}$, respectively.
Thus, under nominal alignment, the residual $\pi$ component does not significantly limit the Bell-state fidelity for our current parameters.}

We must also consider the transmission losses through other optics (mirrors, lenses) $\eta_{\rm trans}=75\%$, the SPCM detector efficiency $\eta_{\rm det}=52\%$, and the pumping and excitation pulse efficiencies $\eta_{\rm pump}=99.1\%$ and $\eta_{\rm exc}=98.3\%$. Combining all these effects, the expected success probability of detecting a single photon is $P_s=1.56\%$.

%% file: main_PRL_version.bbl
\begin{thebibliography}{49}%
\makeatletter
\providecommand \@ifxundefined [1]{%
 \@ifx{#1\undefined}
}%
\providecommand \@ifnum [1]{%
 \ifnum #1\expandafter \@firstoftwo
 \else \expandafter \@secondoftwo
 \fi
}%
\providecommand \@ifx [1]{%
 \ifx #1\expandafter \@firstoftwo
 \else \expandafter \@secondoftwo
 \fi
}%
\providecommand \natexlab [1]{#1}%
\providecommand \enquote  [1]{``#1''}%
\providecommand \bibnamefont  [1]{#1}%
\providecommand \bibfnamefont [1]{#1}%
\providecommand \citenamefont [1]{#1}%
\providecommand \href@noop [0]{\@secondoftwo}%
\providecommand \href [0]{\begingroup \@sanitize@url \@href}%
\providecommand \@href[1]{\@@startlink{#1}\@@href}%
\providecommand \@@href[1]{\endgroup#1\@@endlink}%
\providecommand \@sanitize@url [0]{\catcode `\\12\catcode `\$12\catcode `\&12\catcode `\#12\catcode `\^12\catcode `\_12\catcode `\%12\relax}%
\providecommand \@@startlink[1]{}%
\providecommand \@@endlink[0]{}%
\providecommand \url  [0]{\begingroup\@sanitize@url \@url }%
\providecommand \@url [1]{\endgroup\@href {#1}{\urlprefix }}%
\providecommand \urlprefix  [0]{URL }%
\providecommand \Eprint [0]{\href }%
\providecommand \doibase [0]{https://doi.org/}%
\providecommand \selectlanguage [0]{\@gobble}%
\providecommand \bibinfo  [0]{\@secondoftwo}%
\providecommand \bibfield  [0]{\@secondoftwo}%
\providecommand \translation [1]{[#1]}%
\providecommand \BibitemOpen [0]{}%
\providecommand \bibitemStop [0]{}%
\providecommand \bibitemNoStop [0]{.\EOS\space}%
\providecommand \EOS [0]{\spacefactor3000\relax}%
\providecommand \BibitemShut  [1]{\csname bibitem#1\endcsname}%
\let\auto@bib@innerbib\@empty
%</preamble>
\bibitem [{\citenamefont {Kimble}(2008)}]{Kimble2008}%
  \BibitemOpen
  \bibfield  {author} {\bibinfo {author} {\bibfnamefont {H.~J.}\ \bibnamefont {Kimble}},\ }\bibfield  {title} {\bibinfo {title} {The quantum internet},\ }\href@noop {} {\bibfield  {journal} {\bibinfo  {journal} {Nature}\ }\textbf {\bibinfo {volume} {453}},\ \bibinfo {pages} {1023} (\bibinfo {year} {2008})}\BibitemShut {NoStop}%
\bibitem [{\citenamefont {Wehner}\ \emph {et~al.}(2018)\citenamefont {Wehner}, \citenamefont {Elkouss},\ and\ \citenamefont {Hanson}}]{Wehner2018}%
  \BibitemOpen
  \bibfield  {author} {\bibinfo {author} {\bibfnamefont {S.}~\bibnamefont {Wehner}}, \bibinfo {author} {\bibfnamefont {D.}~\bibnamefont {Elkouss}},\ and\ \bibinfo {author} {\bibfnamefont {R.}~\bibnamefont {Hanson}},\ }\bibfield  {title} {\bibinfo {title} {Quantum internet: a vision for the road ahead},\ }\href@noop {} {\bibfield  {journal} {\bibinfo  {journal} {Science}\ }\textbf {\bibinfo {volume} {362}},\ \bibinfo {pages} {eaam9288} (\bibinfo {year} {2018})}\BibitemShut {NoStop}%
\bibitem [{\citenamefont {Reiserer}\ and\ \citenamefont {Rempe}(2015)}]{Reiserer2015}%
  \BibitemOpen
  \bibfield  {author} {\bibinfo {author} {\bibfnamefont {A.}~\bibnamefont {Reiserer}}\ and\ \bibinfo {author} {\bibfnamefont {G.}~\bibnamefont {Rempe}},\ }\bibfield  {title} {\bibinfo {title} {Cavity-based quantum networks with single atoms and optical photons},\ }\href@noop {} {\bibfield  {journal} {\bibinfo  {journal} {Rev. Mod. Phys.}\ }\textbf {\bibinfo {volume} {87}},\ \bibinfo {pages} {1379} (\bibinfo {year} {2015})}\BibitemShut {NoStop}%
\bibitem [{\citenamefont {Covey}\ \emph {et~al.}(2023)\citenamefont {Covey}, \citenamefont {Weinfurter},\ and\ \citenamefont {Bernien}}]{Covey2023}%
  \BibitemOpen
  \bibfield  {author} {\bibinfo {author} {\bibfnamefont {J.~P.}\ \bibnamefont {Covey}}, \bibinfo {author} {\bibfnamefont {H.}~\bibnamefont {Weinfurter}},\ and\ \bibinfo {author} {\bibfnamefont {H.}~\bibnamefont {Bernien}},\ }\bibfield  {title} {\bibinfo {title} {Quantum networks with neutral atom processing nodes},\ }\href {https://doi.org/10.1038/s41534-023-00759-9} {\bibfield  {journal} {\bibinfo  {journal} {npj Qu. Inf.}\ }\textbf {\bibinfo {volume} {9}},\ \bibinfo {pages} {90} (\bibinfo {year} {2023})}\BibitemShut {NoStop}%
\bibitem [{\citenamefont {Duan}\ \emph {et~al.}(2001)\citenamefont {Duan}, \citenamefont {Lukin}, \citenamefont {Cirac},\ and\ \citenamefont {Zoller}}]{Duan2001}%
  \BibitemOpen
  \bibfield  {author} {\bibinfo {author} {\bibfnamefont {L.~M.}\ \bibnamefont {Duan}}, \bibinfo {author} {\bibfnamefont {M.~D.}\ \bibnamefont {Lukin}}, \bibinfo {author} {\bibfnamefont {J.~I.}\ \bibnamefont {Cirac}},\ and\ \bibinfo {author} {\bibfnamefont {P.}~\bibnamefont {Zoller}},\ }\bibfield  {title} {\bibinfo {title} {Long-distance quantum communication with atomic ensembles and linear optics},\ }\href@noop {} {\bibfield  {journal} {\bibinfo  {journal} {Nature}\ }\textbf {\bibinfo {volume} {414}},\ \bibinfo {pages} {413} (\bibinfo {year} {2001})}\BibitemShut {NoStop}%
\bibitem [{\citenamefont {Sangouard}\ \emph {et~al.}(2011)\citenamefont {Sangouard}, \citenamefont {Simon}, \citenamefont {de~Riedmatten},\ and\ \citenamefont {Gisin}}]{Sangouard2011}%
  \BibitemOpen
  \bibfield  {author} {\bibinfo {author} {\bibfnamefont {N.}~\bibnamefont {Sangouard}}, \bibinfo {author} {\bibfnamefont {C.}~\bibnamefont {Simon}}, \bibinfo {author} {\bibfnamefont {H.}~\bibnamefont {de~Riedmatten}},\ and\ \bibinfo {author} {\bibfnamefont {N.}~\bibnamefont {Gisin}},\ }\bibfield  {title} {\bibinfo {title} {Quantum repeaters based on atomic ensembles and linear optics},\ }\href@noop {} {\bibfield  {journal} {\bibinfo  {journal} {Rev. Mod. Phys.}\ }\textbf {\bibinfo {volume} {83}},\ \bibinfo {pages} {33} (\bibinfo {year} {2011})}\BibitemShut {NoStop}%
\bibitem [{\citenamefont {Hong}\ \emph {et~al.}(1987)\citenamefont {Hong}, \citenamefont {Ou},\ and\ \citenamefont {Mandel}}]{Hong1987}%
  \BibitemOpen
  \bibfield  {author} {\bibinfo {author} {\bibfnamefont {C.~K.}\ \bibnamefont {Hong}}, \bibinfo {author} {\bibfnamefont {Z.~Y.}\ \bibnamefont {Ou}},\ and\ \bibinfo {author} {\bibfnamefont {L.}~\bibnamefont {Mandel}},\ }\bibfield  {title} {\bibinfo {title} {Measurement of subpicosecond time intervals between two photons by interference},\ }\href@noop {} {\bibfield  {journal} {\bibinfo  {journal} {Phys. Rev. Lett.}\ }\textbf {\bibinfo {volume} {59}},\ \bibinfo {pages} {2044} (\bibinfo {year} {1987})}\BibitemShut {NoStop}%
\bibitem [{\citenamefont {Simon}\ and\ \citenamefont {Irvine}(2003)}]{Simon2003}%
  \BibitemOpen
  \bibfield  {author} {\bibinfo {author} {\bibfnamefont {C.}~\bibnamefont {Simon}}\ and\ \bibinfo {author} {\bibfnamefont {W.~T.~M.}\ \bibnamefont {Irvine}},\ }\bibfield  {title} {\bibinfo {title} {Robust long-distance entanglement and a loophole-free {B}ell test with ions and photons},\ }\href {https://doi.org/10.1103/PhysRevLett.91.110405} {\bibfield  {journal} {\bibinfo  {journal} {Phys. Rev. Lett.}\ }\textbf {\bibinfo {volume} {91}},\ \bibinfo {pages} {110405} (\bibinfo {year} {2003})}\BibitemShut {NoStop}%
\bibitem [{\citenamefont {Duan}\ and\ \citenamefont {Kimble}(2003)}]{Duan2003}%
  \BibitemOpen
  \bibfield  {author} {\bibinfo {author} {\bibfnamefont {L.-M.}\ \bibnamefont {Duan}}\ and\ \bibinfo {author} {\bibfnamefont {H.~J.}\ \bibnamefont {Kimble}},\ }\bibfield  {title} {\bibinfo {title} {Efficient engineering of multiatom entanglement through single-photon detections},\ }\href {https://doi.org/10.1103/PhysRevLett.90.253601} {\bibfield  {journal} {\bibinfo  {journal} {Phys. Rev. Lett.}\ }\textbf {\bibinfo {volume} {90}},\ \bibinfo {pages} {253601} (\bibinfo {year} {2003})}\BibitemShut {NoStop}%
\bibitem [{\citenamefont {Browne}\ \emph {et~al.}(2003)\citenamefont {Browne}, \citenamefont {Plenio},\ and\ \citenamefont {Huelga}}]{Browne2003}%
  \BibitemOpen
  \bibfield  {author} {\bibinfo {author} {\bibfnamefont {D.~E.}\ \bibnamefont {Browne}}, \bibinfo {author} {\bibfnamefont {M.~B.}\ \bibnamefont {Plenio}},\ and\ \bibinfo {author} {\bibfnamefont {S.~F.}\ \bibnamefont {Huelga}},\ }\bibfield  {title} {\bibinfo {title} {Robust creation of entanglement between ions in spatially separate cavities},\ }\href {https://doi.org/10.1103/PhysRevLett.91.067901} {\bibfield  {journal} {\bibinfo  {journal} {Phys. Rev. Lett.}\ }\textbf {\bibinfo {volume} {91}},\ \bibinfo {pages} {067901} (\bibinfo {year} {2003})}\BibitemShut {NoStop}%
\bibitem [{\citenamefont {Feng}\ \emph {et~al.}(2003)\citenamefont {Feng}, \citenamefont {Zhang}, \citenamefont {Li}, \citenamefont {Gong},\ and\ \citenamefont {Xu}}]{Feng2003}%
  \BibitemOpen
  \bibfield  {author} {\bibinfo {author} {\bibfnamefont {X.-L.}\ \bibnamefont {Feng}}, \bibinfo {author} {\bibfnamefont {Z.-M.}\ \bibnamefont {Zhang}}, \bibinfo {author} {\bibfnamefont {X.-D.}\ \bibnamefont {Li}}, \bibinfo {author} {\bibfnamefont {S.-Q.}\ \bibnamefont {Gong}},\ and\ \bibinfo {author} {\bibfnamefont {Z.-Z.}\ \bibnamefont {Xu}},\ }\bibfield  {title} {\bibinfo {title} {Entangling distant atoms by interference of polarized photons},\ }\href {https://doi.org/10.1103/PhysRevLett.90.217902} {\bibfield  {journal} {\bibinfo  {journal} {Phys. Rev. Lett.}\ }\textbf {\bibinfo {volume} {90}},\ \bibinfo {pages} {217902} (\bibinfo {year} {2003})}\BibitemShut {NoStop}%
\bibitem [{\citenamefont {Volz}\ \emph {et~al.}(2006)\citenamefont {Volz}, \citenamefont {Weber}, \citenamefont {Schlenk}, \citenamefont {Rosenfeld}, \citenamefont {Vrana}, \citenamefont {Saucke}, \citenamefont {Kurtsiefer},\ and\ \citenamefont {Weinfurter}}]{Volz2006}%
  \BibitemOpen
  \bibfield  {author} {\bibinfo {author} {\bibfnamefont {J.}~\bibnamefont {Volz}}, \bibinfo {author} {\bibfnamefont {M.}~\bibnamefont {Weber}}, \bibinfo {author} {\bibfnamefont {D.}~\bibnamefont {Schlenk}}, \bibinfo {author} {\bibfnamefont {W.}~\bibnamefont {Rosenfeld}}, \bibinfo {author} {\bibfnamefont {J.}~\bibnamefont {Vrana}}, \bibinfo {author} {\bibfnamefont {K.}~\bibnamefont {Saucke}}, \bibinfo {author} {\bibfnamefont {C.}~\bibnamefont {Kurtsiefer}},\ and\ \bibinfo {author} {\bibfnamefont {H.}~\bibnamefont {Weinfurter}},\ }\bibfield  {title} {\bibinfo {title} {Observation of entanglement of a single photon with a trapped atom},\ }\href@noop {} {\bibfield  {journal} {\bibinfo  {journal} {Phys. Rev. Lett.}\ }\textbf {\bibinfo {volume} {96}},\ \bibinfo {pages} {030404} (\bibinfo {year} {2006})}\BibitemShut {NoStop}%
\bibitem [{\citenamefont {Wilk}\ \emph {et~al.}(2007{\natexlab{a}})\citenamefont {Wilk}, \citenamefont {Webster}, \citenamefont {Kuhn},\ and\ \citenamefont {Rempe}}]{Wilk2007b}%
  \BibitemOpen
  \bibfield  {author} {\bibinfo {author} {\bibfnamefont {T.}~\bibnamefont {Wilk}}, \bibinfo {author} {\bibfnamefont {S.~C.}\ \bibnamefont {Webster}}, \bibinfo {author} {\bibfnamefont {A.}~\bibnamefont {Kuhn}},\ and\ \bibinfo {author} {\bibfnamefont {G.}~\bibnamefont {Rempe}},\ }\bibfield  {title} {\bibinfo {title} {Single-atom single-photon quantum interface},\ }\href@noop {} {\bibfield  {journal} {\bibinfo  {journal} {Science}\ }\textbf {\bibinfo {volume} {317}},\ \bibinfo {pages} {488} (\bibinfo {year} {2007}{\natexlab{a}})}\BibitemShut {NoStop}%
\bibitem [{\citenamefont {van Leent}\ \emph {et~al.}(2020)\citenamefont {van Leent}, \citenamefont {Bock}, \citenamefont {Garthoff}, \citenamefont {Redeker}, \citenamefont {Zhang}, \citenamefont {Bauer}, \citenamefont {Rosenfeld}, \citenamefont {Becher},\ and\ \citenamefont {Weinfurter}}]{vanLeent2020}%
  \BibitemOpen
  \bibfield  {author} {\bibinfo {author} {\bibfnamefont {T.}~\bibnamefont {van Leent}}, \bibinfo {author} {\bibfnamefont {M.}~\bibnamefont {Bock}}, \bibinfo {author} {\bibfnamefont {R.}~\bibnamefont {Garthoff}}, \bibinfo {author} {\bibfnamefont {K.}~\bibnamefont {Redeker}}, \bibinfo {author} {\bibfnamefont {W.}~\bibnamefont {Zhang}}, \bibinfo {author} {\bibfnamefont {T.}~\bibnamefont {Bauer}}, \bibinfo {author} {\bibfnamefont {W.}~\bibnamefont {Rosenfeld}}, \bibinfo {author} {\bibfnamefont {C.}~\bibnamefont {Becher}},\ and\ \bibinfo {author} {\bibfnamefont {H.}~\bibnamefont {Weinfurter}},\ }\bibfield  {title} {\bibinfo {title} {Long-distance distribution of atom-photon entanglement at telecom wavelength},\ }\href@noop {} {\bibfield  {journal} {\bibinfo  {journal} {Phys. Rev. Lett.}\ }\textbf {\bibinfo {volume} {124}},\ \bibinfo {pages} {010510} (\bibinfo {year} {2020})}\BibitemShut {NoStop}%
\bibitem [{\citenamefont {Zhang}\ \emph {et~al.}(2022)\citenamefont {Zhang}, \citenamefont {van Leent}, \citenamefont {Redeker}, \citenamefont {Garthoff}, \citenamefont {Schwonnek}, \citenamefont {Fertig}, \citenamefont {Eppelt}, \citenamefont {Rosenfeld}, \citenamefont {Scarani}, \citenamefont {Lim},\ and\ \citenamefont {Weinfurter}}]{Zhang2022}%
  \BibitemOpen
  \bibfield  {author} {\bibinfo {author} {\bibfnamefont {W.}~\bibnamefont {Zhang}}, \bibinfo {author} {\bibfnamefont {T.}~\bibnamefont {van Leent}}, \bibinfo {author} {\bibfnamefont {K.}~\bibnamefont {Redeker}}, \bibinfo {author} {\bibfnamefont {R.}~\bibnamefont {Garthoff}}, \bibinfo {author} {\bibfnamefont {R.}~\bibnamefont {Schwonnek}}, \bibinfo {author} {\bibfnamefont {F.}~\bibnamefont {Fertig}}, \bibinfo {author} {\bibfnamefont {S.}~\bibnamefont {Eppelt}}, \bibinfo {author} {\bibfnamefont {W.}~\bibnamefont {Rosenfeld}}, \bibinfo {author} {\bibfnamefont {V.}~\bibnamefont {Scarani}}, \bibinfo {author} {\bibfnamefont {C.~C.-W.}\ \bibnamefont {Lim}},\ and\ \bibinfo {author} {\bibfnamefont {H.}~\bibnamefont {Weinfurter}},\ }\bibfield  {title} {\bibinfo {title} {A device-independent quantum key distribution system for distant users},\ }\href {https://doi.org/10.1038/s41586-022-04891-y} {\bibfield  {journal} {\bibinfo  {journal} {Nature}\ }\textbf {\bibinfo {volume} {607}},\ \bibinfo {pages} {687} (\bibinfo
  {year} {2022})}\BibitemShut {NoStop}%
\bibitem [{\citenamefont {Hartung}\ \emph {et~al.}(2024)\citenamefont {Hartung}, \citenamefont {Seubert}, \citenamefont {Welte}, \citenamefont {Distante},\ and\ \citenamefont {Rempe}}]{Hartung2024}%
  \BibitemOpen
  \bibfield  {author} {\bibinfo {author} {\bibfnamefont {L.}~\bibnamefont {Hartung}}, \bibinfo {author} {\bibfnamefont {M.}~\bibnamefont {Seubert}}, \bibinfo {author} {\bibfnamefont {S.}~\bibnamefont {Welte}}, \bibinfo {author} {\bibfnamefont {E.}~\bibnamefont {Distante}},\ and\ \bibinfo {author} {\bibfnamefont {G.}~\bibnamefont {Rempe}},\ }\bibfield  {title} {\bibinfo {title} {A quantum-network register assembled with optical tweezers in an optical cavity},\ }\href@noop {} {\bibfield  {journal} {\bibinfo  {journal} {Science}\ }\textbf {\bibinfo {volume} {385}},\ \bibinfo {pages} {179} (\bibinfo {year} {2024})}\BibitemShut {NoStop}%
\bibitem [{\citenamefont {Li}\ \emph {et~al.}(2025)\citenamefont {Li}, \citenamefont {Hu}, \citenamefont {Jia}, \citenamefont {Huie}, \citenamefont {Sun}, \citenamefont {Aakash}, \citenamefont {Dong}, \citenamefont {Hiri-O-Tuppa},\ and\ \citenamefont {Covey}}]{LLi2025}%
  \BibitemOpen
  \bibfield  {author} {\bibinfo {author} {\bibfnamefont {L.}~\bibnamefont {Li}}, \bibinfo {author} {\bibfnamefont {X.}~\bibnamefont {Hu}}, \bibinfo {author} {\bibfnamefont {Z.}~\bibnamefont {Jia}}, \bibinfo {author} {\bibfnamefont {W.}~\bibnamefont {Huie}}, \bibinfo {author} {\bibfnamefont {W.~K.~C.}\ \bibnamefont {Sun}}, \bibinfo {author} {\bibnamefont {Aakash}}, \bibinfo {author} {\bibfnamefont {Y.}~\bibnamefont {Dong}}, \bibinfo {author} {\bibfnamefont {N.}~\bibnamefont {Hiri-O-Tuppa}},\ and\ \bibinfo {author} {\bibfnamefont {J.~P.}\ \bibnamefont {Covey}},\ }\bibfield  {title} {\bibinfo {title} {Parallelized telecom quantum networking with an ytterbium-171 atom array},\ }\href {https://doi.org/10.1038/s41567-025-03022-4} {\bibfield  {journal} {\bibinfo  {journal} {Nat. Phys.}\ }\textbf {\bibinfo {volume} {21}},\ \bibinfo {pages} {1826} (\bibinfo {year} {2025})}\BibitemShut {NoStop}%
\bibitem [{\citenamefont {Safari}\ \emph {et~al.}(2026)\citenamefont {Safari}, \citenamefont {Oh}, \citenamefont {Huft}, \citenamefont {Chase}, \citenamefont {Zhang},\ and\ \citenamefont {Saffman}}]{Safari2026}%
  \BibitemOpen
  \bibfield  {author} {\bibinfo {author} {\bibfnamefont {A.}~\bibnamefont {Safari}}, \bibinfo {author} {\bibfnamefont {E.}~\bibnamefont {Oh}}, \bibinfo {author} {\bibfnamefont {P.}~\bibnamefont {Huft}}, \bibinfo {author} {\bibfnamefont {G.}~\bibnamefont {Chase}}, \bibinfo {author} {\bibfnamefont {J.}~\bibnamefont {Zhang}},\ and\ \bibinfo {author} {\bibfnamefont {M.}~\bibnamefont {Saffman}},\ }\bibfield  {title} {\bibinfo {title} {Efficient and compact quantum network node based on a parabolic mirror on an optical chip},\ }\href {https://doi.org/10.1103/fhf3-3nzb} {\bibfield  {journal} {\bibinfo  {journal} {PRX Quantum}\ }\textbf {\bibinfo {volume} {7}},\ \bibinfo {pages} {033008} (\bibinfo {year} {2026})}\BibitemShut {NoStop}%
\bibitem [{\citenamefont {Stephenson}\ \emph {et~al.}(2020)\citenamefont {Stephenson}, \citenamefont {Nadlinger}, \citenamefont {Nichol}, \citenamefont {An}, \citenamefont {Drmota}, \citenamefont {Ballance}, \citenamefont {Thirumalai}, \citenamefont {Goodwin}, \citenamefont {Lucas},\ and\ \citenamefont {Ballance}}]{Stephenson2020}%
  \BibitemOpen
  \bibfield  {author} {\bibinfo {author} {\bibfnamefont {L.~J.}\ \bibnamefont {Stephenson}}, \bibinfo {author} {\bibfnamefont {D.~P.}\ \bibnamefont {Nadlinger}}, \bibinfo {author} {\bibfnamefont {B.~C.}\ \bibnamefont {Nichol}}, \bibinfo {author} {\bibfnamefont {S.}~\bibnamefont {An}}, \bibinfo {author} {\bibfnamefont {P.}~\bibnamefont {Drmota}}, \bibinfo {author} {\bibfnamefont {T.~G.}\ \bibnamefont {Ballance}}, \bibinfo {author} {\bibfnamefont {K.}~\bibnamefont {Thirumalai}}, \bibinfo {author} {\bibfnamefont {J.~F.}\ \bibnamefont {Goodwin}}, \bibinfo {author} {\bibfnamefont {D.~M.}\ \bibnamefont {Lucas}},\ and\ \bibinfo {author} {\bibfnamefont {C.~J.}\ \bibnamefont {Ballance}},\ }\bibfield  {title} {\bibinfo {title} {High-rate, high-fidelity entanglement of qubits across an elementary quantum network},\ }\href@noop {} {\bibfield  {journal} {\bibinfo  {journal} {Phys. Rev. Lett.}\ }\textbf {\bibinfo {volume} {124}},\ \bibinfo {pages} {110501} (\bibinfo {year} {2020})}\BibitemShut {NoStop}%
\bibitem [{\citenamefont {Krutyanskiy}\ \emph {et~al.}(2023)\citenamefont {Krutyanskiy}, \citenamefont {Galli}, \citenamefont {Krcmarsky}, \citenamefont {Baier}, \citenamefont {Fioretto}, \citenamefont {Pu}, \citenamefont {Mazloom}, \citenamefont {Sekatski}, \citenamefont {Canteri}, \citenamefont {Teller}, \citenamefont {Schupp}, \citenamefont {Bate}, \citenamefont {Meraner}, \citenamefont {Sangouard}, \citenamefont {Lanyon},\ and\ \citenamefont {Northup}}]{Krutyanskiy2023}%
  \BibitemOpen
  \bibfield  {author} {\bibinfo {author} {\bibfnamefont {V.}~\bibnamefont {Krutyanskiy}}, \bibinfo {author} {\bibfnamefont {M.}~\bibnamefont {Galli}}, \bibinfo {author} {\bibfnamefont {V.}~\bibnamefont {Krcmarsky}}, \bibinfo {author} {\bibfnamefont {S.}~\bibnamefont {Baier}}, \bibinfo {author} {\bibfnamefont {D.~A.}\ \bibnamefont {Fioretto}}, \bibinfo {author} {\bibfnamefont {Y.}~\bibnamefont {Pu}}, \bibinfo {author} {\bibfnamefont {A.}~\bibnamefont {Mazloom}}, \bibinfo {author} {\bibfnamefont {P.}~\bibnamefont {Sekatski}}, \bibinfo {author} {\bibfnamefont {M.}~\bibnamefont {Canteri}}, \bibinfo {author} {\bibfnamefont {M.}~\bibnamefont {Teller}}, \bibinfo {author} {\bibfnamefont {J.}~\bibnamefont {Schupp}}, \bibinfo {author} {\bibfnamefont {J.}~\bibnamefont {Bate}}, \bibinfo {author} {\bibfnamefont {M.}~\bibnamefont {Meraner}}, \bibinfo {author} {\bibfnamefont {N.}~\bibnamefont {Sangouard}}, \bibinfo {author} {\bibfnamefont {B.~P.}\ \bibnamefont {Lanyon}},\ and\ \bibinfo {author} {\bibfnamefont {T.~E.}\
  \bibnamefont {Northup}},\ }\bibfield  {title} {\bibinfo {title} {Entanglement of trapped-ion qubits separated by 230 meters},\ }\href {https://doi.org/10.1103/PhysRevLett.130.050803} {\bibfield  {journal} {\bibinfo  {journal} {Phys. Rev. Lett.}\ }\textbf {\bibinfo {volume} {130}},\ \bibinfo {pages} {050803} (\bibinfo {year} {2023})}\BibitemShut {NoStop}%
\bibitem [{\citenamefont {O'Reilly}\ \emph {et~al.}(2024)\citenamefont {O'Reilly}, \citenamefont {Toh}, \citenamefont {Goetting}, \citenamefont {Saha}, \citenamefont {Shalaev}, \citenamefont {Carter}, \citenamefont {Risinger}, \citenamefont {Kalakuntla}, \citenamefont {Li}, \citenamefont {Verma},\ and\ \citenamefont {Monroe}}]{OReilly2025}%
  \BibitemOpen
  \bibfield  {author} {\bibinfo {author} {\bibfnamefont {J.}~\bibnamefont {O'Reilly}}, \bibinfo {author} {\bibfnamefont {G.}~\bibnamefont {Toh}}, \bibinfo {author} {\bibfnamefont {I.}~\bibnamefont {Goetting}}, \bibinfo {author} {\bibfnamefont {S.}~\bibnamefont {Saha}}, \bibinfo {author} {\bibfnamefont {M.}~\bibnamefont {Shalaev}}, \bibinfo {author} {\bibfnamefont {A.~L.}\ \bibnamefont {Carter}}, \bibinfo {author} {\bibfnamefont {A.}~\bibnamefont {Risinger}}, \bibinfo {author} {\bibfnamefont {A.}~\bibnamefont {Kalakuntla}}, \bibinfo {author} {\bibfnamefont {T.}~\bibnamefont {Li}}, \bibinfo {author} {\bibfnamefont {A.}~\bibnamefont {Verma}},\ and\ \bibinfo {author} {\bibfnamefont {C.}~\bibnamefont {Monroe}},\ }\bibfield  {title} {\bibinfo {title} {Fast photon-mediated entanglement of continuously cooled trapped ions for quantum networking},\ }\href@noop {} {\bibfield  {journal} {\bibinfo  {journal} {Phys. Rev. Lett.}\ }\textbf {\bibinfo {volume} {133}},\ \bibinfo {pages} {090802} (\bibinfo {year}
  {2024})}\BibitemShut {NoStop}%
\bibitem [{\citenamefont {Pompili}\ \emph {et~al.}(2021)\citenamefont {Pompili}, \citenamefont {Hermans}, \citenamefont {Baier}, \citenamefont {Beukers}, \citenamefont {Humphreys}, \citenamefont {Schouten}, \citenamefont {Vermeulen}, \citenamefont {Tiggelman}, \citenamefont {dos Santos~Martins}, \citenamefont {Dirkse}, \citenamefont {Wehner},\ and\ \citenamefont {Hanson}}]{Pompili2021}%
  \BibitemOpen
  \bibfield  {author} {\bibinfo {author} {\bibfnamefont {M.}~\bibnamefont {Pompili}}, \bibinfo {author} {\bibfnamefont {S.~L.~N.}\ \bibnamefont {Hermans}}, \bibinfo {author} {\bibfnamefont {S.}~\bibnamefont {Baier}}, \bibinfo {author} {\bibfnamefont {H.~K.~C.}\ \bibnamefont {Beukers}}, \bibinfo {author} {\bibfnamefont {P.~C.}\ \bibnamefont {Humphreys}}, \bibinfo {author} {\bibfnamefont {R.~N.}\ \bibnamefont {Schouten}}, \bibinfo {author} {\bibfnamefont {R.~F.~L.}\ \bibnamefont {Vermeulen}}, \bibinfo {author} {\bibfnamefont {M.~J.}\ \bibnamefont {Tiggelman}}, \bibinfo {author} {\bibfnamefont {L.}~\bibnamefont {dos Santos~Martins}}, \bibinfo {author} {\bibfnamefont {B.}~\bibnamefont {Dirkse}}, \bibinfo {author} {\bibfnamefont {S.}~\bibnamefont {Wehner}},\ and\ \bibinfo {author} {\bibfnamefont {R.}~\bibnamefont {Hanson}},\ }\bibfield  {title} {\bibinfo {title} {Realization of a multinode quantum network of remote solid-state qubits},\ }\href@noop {} {\bibfield  {journal} {\bibinfo  {journal} {Science}\ }\textbf
  {\bibinfo {volume} {372}},\ \bibinfo {pages} {259} (\bibinfo {year} {2021})}\BibitemShut {NoStop}%
\bibitem [{\citenamefont {Stockill}\ \emph {et~al.}(2017)\citenamefont {Stockill}, \citenamefont {Stanley}, \citenamefont {Huthmacher}, \citenamefont {Clarke}, \citenamefont {Hugues}, \citenamefont {Miller}, \citenamefont {Matthiesen}, \citenamefont {Le~Gall},\ and\ \citenamefont {Atat\"ure}}]{Stockill2017}%
  \BibitemOpen
  \bibfield  {author} {\bibinfo {author} {\bibfnamefont {R.}~\bibnamefont {Stockill}}, \bibinfo {author} {\bibfnamefont {M.~J.}\ \bibnamefont {Stanley}}, \bibinfo {author} {\bibfnamefont {L.}~\bibnamefont {Huthmacher}}, \bibinfo {author} {\bibfnamefont {E.}~\bibnamefont {Clarke}}, \bibinfo {author} {\bibfnamefont {M.}~\bibnamefont {Hugues}}, \bibinfo {author} {\bibfnamefont {A.~J.}\ \bibnamefont {Miller}}, \bibinfo {author} {\bibfnamefont {C.}~\bibnamefont {Matthiesen}}, \bibinfo {author} {\bibfnamefont {C.}~\bibnamefont {Le~Gall}},\ and\ \bibinfo {author} {\bibfnamefont {M.}~\bibnamefont {Atat\"ure}},\ }\bibfield  {title} {\bibinfo {title} {Phase-tuned entangled state generation between distant spin qubits},\ }\href {https://doi.org/10.1103/PhysRevLett.119.010503} {\bibfield  {journal} {\bibinfo  {journal} {Phys. Rev. Lett.}\ }\textbf {\bibinfo {volume} {119}},\ \bibinfo {pages} {010503} (\bibinfo {year} {2017})}\BibitemShut {NoStop}%
\bibitem [{\citenamefont {Welte}\ \emph {et~al.}(2021)\citenamefont {Welte}, \citenamefont {Thomas}, \citenamefont {Hartung}, \citenamefont {Daiss}, \citenamefont {Langenfeld}, \citenamefont {Morin}, \citenamefont {Rempe},\ and\ \citenamefont {Distante}}]{Welte2021}%
  \BibitemOpen
  \bibfield  {author} {\bibinfo {author} {\bibfnamefont {S.}~\bibnamefont {Welte}}, \bibinfo {author} {\bibfnamefont {P.}~\bibnamefont {Thomas}}, \bibinfo {author} {\bibfnamefont {L.}~\bibnamefont {Hartung}}, \bibinfo {author} {\bibfnamefont {S.}~\bibnamefont {Daiss}}, \bibinfo {author} {\bibfnamefont {S.}~\bibnamefont {Langenfeld}}, \bibinfo {author} {\bibfnamefont {O.}~\bibnamefont {Morin}}, \bibinfo {author} {\bibfnamefont {G.}~\bibnamefont {Rempe}},\ and\ \bibinfo {author} {\bibfnamefont {E.}~\bibnamefont {Distante}},\ }\bibfield  {title} {\bibinfo {title} {A nondestructive {B}ell-state measurement on two distant atomic qubits},\ }\href {https://doi.org/10.1038/s41566-021-00802-1} {\bibfield  {journal} {\bibinfo  {journal} {Nat. Photon.}\ }\textbf {\bibinfo {volume} {15}},\ \bibinfo {pages} {504} (\bibinfo {year} {2021})}\BibitemShut {NoStop}%
\bibitem [{\citenamefont {van Leent}\ \emph {et~al.}(2022)\citenamefont {van Leent}, \citenamefont {Bock}, \citenamefont {Fertig}, \citenamefont {Garthoff}, \citenamefont {Eppelt}, \citenamefont {Zhou}, \citenamefont {Malik}, \citenamefont {Seubert}, \citenamefont {Bauer}, \citenamefont {Rosenfeld}, \citenamefont {Zhang}, \citenamefont {Becher},\ and\ \citenamefont {Weinfurter}}]{vanLeent2022}%
  \BibitemOpen
  \bibfield  {author} {\bibinfo {author} {\bibfnamefont {T.}~\bibnamefont {van Leent}}, \bibinfo {author} {\bibfnamefont {M.}~\bibnamefont {Bock}}, \bibinfo {author} {\bibfnamefont {F.}~\bibnamefont {Fertig}}, \bibinfo {author} {\bibfnamefont {R.}~\bibnamefont {Garthoff}}, \bibinfo {author} {\bibfnamefont {S.}~\bibnamefont {Eppelt}}, \bibinfo {author} {\bibfnamefont {Y.}~\bibnamefont {Zhou}}, \bibinfo {author} {\bibfnamefont {P.}~\bibnamefont {Malik}}, \bibinfo {author} {\bibfnamefont {M.}~\bibnamefont {Seubert}}, \bibinfo {author} {\bibfnamefont {T.}~\bibnamefont {Bauer}}, \bibinfo {author} {\bibfnamefont {W.}~\bibnamefont {Rosenfeld}}, \bibinfo {author} {\bibfnamefont {W.}~\bibnamefont {Zhang}}, \bibinfo {author} {\bibfnamefont {C.}~\bibnamefont {Becher}},\ and\ \bibinfo {author} {\bibfnamefont {H.}~\bibnamefont {Weinfurter}},\ }\bibfield  {title} {\bibinfo {title} {Entangling single atoms over 33 km telecom fibre},\ }\href@noop {} {\bibfield  {journal} {\bibinfo  {journal} {Nature}\ }\textbf {\bibinfo
  {volume} {607}},\ \bibinfo {pages} {69} (\bibinfo {year} {2022})}\BibitemShut {NoStop}%
\bibitem [{\citenamefont {Barredo}\ \emph {et~al.}(2016)\citenamefont {Barredo}, \citenamefont {de~Les\'el\'euc}, \citenamefont {Lienhard}, \citenamefont {Lahaye},\ and\ \citenamefont {Browaeys}}]{Barredo2016}%
  \BibitemOpen
  \bibfield  {author} {\bibinfo {author} {\bibfnamefont {D.}~\bibnamefont {Barredo}}, \bibinfo {author} {\bibfnamefont {S.}~\bibnamefont {de~Les\'el\'euc}}, \bibinfo {author} {\bibfnamefont {V.}~\bibnamefont {Lienhard}}, \bibinfo {author} {\bibfnamefont {T.}~\bibnamefont {Lahaye}},\ and\ \bibinfo {author} {\bibfnamefont {A.}~\bibnamefont {Browaeys}},\ }\bibfield  {title} {\bibinfo {title} {An atom-by-atom assembler of defect-free arbitrary two-dimensional atomic arrays},\ }\href@noop {} {\bibfield  {journal} {\bibinfo  {journal} {Science}\ }\textbf {\bibinfo {volume} {354}},\ \bibinfo {pages} {1021} (\bibinfo {year} {2016})}\BibitemShut {NoStop}%
\bibitem [{\citenamefont {Endres}\ \emph {et~al.}(2016)\citenamefont {Endres}, \citenamefont {Bernien}, \citenamefont {Keesling}, \citenamefont {Levine}, \citenamefont {Anschuetz}, \citenamefont {Krajenbrink}, \citenamefont {Senko}, \citenamefont {Vuletic}, \citenamefont {Greiner},\ and\ \citenamefont {Lukin}}]{Endres2016}%
  \BibitemOpen
  \bibfield  {author} {\bibinfo {author} {\bibfnamefont {M.}~\bibnamefont {Endres}}, \bibinfo {author} {\bibfnamefont {H.}~\bibnamefont {Bernien}}, \bibinfo {author} {\bibfnamefont {A.}~\bibnamefont {Keesling}}, \bibinfo {author} {\bibfnamefont {H.}~\bibnamefont {Levine}}, \bibinfo {author} {\bibfnamefont {E.~R.}\ \bibnamefont {Anschuetz}}, \bibinfo {author} {\bibfnamefont {A.}~\bibnamefont {Krajenbrink}}, \bibinfo {author} {\bibfnamefont {C.}~\bibnamefont {Senko}}, \bibinfo {author} {\bibfnamefont {V.}~\bibnamefont {Vuletic}}, \bibinfo {author} {\bibfnamefont {M.}~\bibnamefont {Greiner}},\ and\ \bibinfo {author} {\bibfnamefont {M.~D.}\ \bibnamefont {Lukin}},\ }\bibfield  {title} {\bibinfo {title} {Atom-by-atom assembly of defect-free one-dimensional cold atom arrays},\ }\href@noop {} {\bibfield  {journal} {\bibinfo  {journal} {Science}\ }\textbf {\bibinfo {volume} {354}},\ \bibinfo {pages} {1024} (\bibinfo {year} {2016})}\BibitemShut {NoStop}%
\bibitem [{\citenamefont {Muniz}\ \emph {et~al.}(2025)\citenamefont {Muniz}, \citenamefont {Stone}, \citenamefont {Stack}, \citenamefont {Jaffe}, \citenamefont {Kindem}, \citenamefont {Wadleigh}, \citenamefont {Zalys-Geller}, \citenamefont {Zhang}, \citenamefont {Chen}, \citenamefont {Norcia}, \citenamefont {Epstein}, \citenamefont {Halperin}, \citenamefont {Hummel}, \citenamefont {Wilkason}, \citenamefont {Li}, \citenamefont {Barnes}, \citenamefont {Battaglino}, \citenamefont {Bohdanowicz}, \citenamefont {Booth}, \citenamefont {Brown}, \citenamefont {Brown}, \citenamefont {Cairncross}, \citenamefont {Cassella}, \citenamefont {Coxe}, \citenamefont {Crow}, \citenamefont {Feldkamp}, \citenamefont {Griger}, \citenamefont {Heinz}, \citenamefont {Jones}, \citenamefont {Kim}, \citenamefont {King}, \citenamefont {Kotru}, \citenamefont {Lauigan}, \citenamefont {Marjanovic}, \citenamefont {Megidish}, \citenamefont {Meredith}, \citenamefont {McDonald}, \citenamefont {Morshead}, \citenamefont {Narayanaswami},
  \citenamefont {Nishiguchi}, \citenamefont {Paule}, \citenamefont {Pawlak}, \citenamefont {Pudenz}, \citenamefont {P\'erez}, \citenamefont {Ryou}, \citenamefont {Simon}, \citenamefont {Smull}, \citenamefont {Urbanek}, \citenamefont {van~de Veerdonk}, \citenamefont {Vendeiro}, \citenamefont {Wu}, \citenamefont {Xie},\ and\ \citenamefont {Bloom}}]{Muniz2025}%
  \BibitemOpen
  \bibfield  {author} {\bibinfo {author} {\bibfnamefont {J.~A.}\ \bibnamefont {Muniz}}, \bibinfo {author} {\bibfnamefont {M.}~\bibnamefont {Stone}}, \bibinfo {author} {\bibfnamefont {D.~T.}\ \bibnamefont {Stack}}, \bibinfo {author} {\bibfnamefont {M.}~\bibnamefont {Jaffe}}, \bibinfo {author} {\bibfnamefont {J.~M.}\ \bibnamefont {Kindem}}, \bibinfo {author} {\bibfnamefont {L.}~\bibnamefont {Wadleigh}}, \bibinfo {author} {\bibfnamefont {E.}~\bibnamefont {Zalys-Geller}}, \bibinfo {author} {\bibfnamefont {X.}~\bibnamefont {Zhang}}, \bibinfo {author} {\bibfnamefont {C.-A.}\ \bibnamefont {Chen}}, \bibinfo {author} {\bibfnamefont {M.~A.}\ \bibnamefont {Norcia}}, \bibinfo {author} {\bibfnamefont {J.}~\bibnamefont {Epstein}}, \bibinfo {author} {\bibfnamefont {E.}~\bibnamefont {Halperin}}, \bibinfo {author} {\bibfnamefont {F.}~\bibnamefont {Hummel}}, \bibinfo {author} {\bibfnamefont {T.}~\bibnamefont {Wilkason}}, \bibinfo {author} {\bibfnamefont {M.}~\bibnamefont {Li}}, \bibinfo {author} {\bibfnamefont
  {K.}~\bibnamefont {Barnes}}, \bibinfo {author} {\bibfnamefont {P.}~\bibnamefont {Battaglino}}, \bibinfo {author} {\bibfnamefont {T.~C.}\ \bibnamefont {Bohdanowicz}}, \bibinfo {author} {\bibfnamefont {G.}~\bibnamefont {Booth}}, \bibinfo {author} {\bibfnamefont {A.}~\bibnamefont {Brown}}, \bibinfo {author} {\bibfnamefont {M.~O.}\ \bibnamefont {Brown}}, \bibinfo {author} {\bibfnamefont {W.~B.}\ \bibnamefont {Cairncross}}, \bibinfo {author} {\bibfnamefont {K.}~\bibnamefont {Cassella}}, \bibinfo {author} {\bibfnamefont {R.}~\bibnamefont {Coxe}}, \bibinfo {author} {\bibfnamefont {D.}~\bibnamefont {Crow}}, \bibinfo {author} {\bibfnamefont {M.}~\bibnamefont {Feldkamp}}, \bibinfo {author} {\bibfnamefont {C.}~\bibnamefont {Griger}}, \bibinfo {author} {\bibfnamefont {A.}~\bibnamefont {Heinz}}, \bibinfo {author} {\bibfnamefont {A.~M.~W.}\ \bibnamefont {Jones}}, \bibinfo {author} {\bibfnamefont {H.}~\bibnamefont {Kim}}, \bibinfo {author} {\bibfnamefont {J.}~\bibnamefont {King}}, \bibinfo {author} {\bibfnamefont
  {K.}~\bibnamefont {Kotru}}, \bibinfo {author} {\bibfnamefont {J.}~\bibnamefont {Lauigan}}, \bibinfo {author} {\bibfnamefont {J.}~\bibnamefont {Marjanovic}}, \bibinfo {author} {\bibfnamefont {E.}~\bibnamefont {Megidish}}, \bibinfo {author} {\bibfnamefont {M.}~\bibnamefont {Meredith}}, \bibinfo {author} {\bibfnamefont {M.}~\bibnamefont {McDonald}}, \bibinfo {author} {\bibfnamefont {R.}~\bibnamefont {Morshead}}, \bibinfo {author} {\bibfnamefont {S.}~\bibnamefont {Narayanaswami}}, \bibinfo {author} {\bibfnamefont {C.}~\bibnamefont {Nishiguchi}}, \bibinfo {author} {\bibfnamefont {T.}~\bibnamefont {Paule}}, \bibinfo {author} {\bibfnamefont {K.~A.}\ \bibnamefont {Pawlak}}, \bibinfo {author} {\bibfnamefont {K.~L.}\ \bibnamefont {Pudenz}}, \bibinfo {author} {\bibfnamefont {D.~R.}\ \bibnamefont {P\'erez}}, \bibinfo {author} {\bibfnamefont {A.}~\bibnamefont {Ryou}}, \bibinfo {author} {\bibfnamefont {J.}~\bibnamefont {Simon}}, \bibinfo {author} {\bibfnamefont {A.}~\bibnamefont {Smull}}, \bibinfo {author} {\bibfnamefont
  {M.}~\bibnamefont {Urbanek}}, \bibinfo {author} {\bibfnamefont {R.~J.~M.}\ \bibnamefont {van~de Veerdonk}}, \bibinfo {author} {\bibfnamefont {Z.}~\bibnamefont {Vendeiro}}, \bibinfo {author} {\bibfnamefont {T.-Y.}\ \bibnamefont {Wu}}, \bibinfo {author} {\bibfnamefont {X.}~\bibnamefont {Xie}},\ and\ \bibinfo {author} {\bibfnamefont {B.~J.}\ \bibnamefont {Bloom}},\ }\bibfield  {title} {\bibinfo {title} {High-fidelity universal gates in the ${}^{171}$$\mathrm{Yb}$ ground-state nuclear-spin qubit},\ }\href@noop {} {\bibfield  {journal} {\bibinfo  {journal} {PRX Quantum}\ }\textbf {\bibinfo {volume} {6}},\ \bibinfo {pages} {020334} (\bibinfo {year} {2025})}\BibitemShut {NoStop}%
\bibitem [{\citenamefont {Bluvstein}\ \emph {et~al.}(2026)\citenamefont {Bluvstein}, \citenamefont {Geim}, \citenamefont {Li}, \citenamefont {Evered}, \citenamefont {Bonilla~Ataides}, \citenamefont {Baranes}, \citenamefont {Gu}, \citenamefont {Manovitz}, \citenamefont {Xu}, \citenamefont {Kalinowski}, \citenamefont {Majidy}, \citenamefont {Kokail}, \citenamefont {Maskara}, \citenamefont {Trapp}, \citenamefont {Stewart}, \citenamefont {Hollerith}, \citenamefont {Zhou}, \citenamefont {Gullans}, \citenamefont {Yelin}, \citenamefont {Greiner}, \citenamefont {Vuletić}, \citenamefont {Cain},\ and\ \citenamefont {Lukin}}]{Bluvstein2026}%
  \BibitemOpen
  \bibfield  {author} {\bibinfo {author} {\bibfnamefont {D.}~\bibnamefont {Bluvstein}}, \bibinfo {author} {\bibfnamefont {A.~A.}\ \bibnamefont {Geim}}, \bibinfo {author} {\bibfnamefont {S.~H.}\ \bibnamefont {Li}}, \bibinfo {author} {\bibfnamefont {S.~J.}\ \bibnamefont {Evered}}, \bibinfo {author} {\bibfnamefont {J.~P.}\ \bibnamefont {Bonilla~Ataides}}, \bibinfo {author} {\bibfnamefont {G.}~\bibnamefont {Baranes}}, \bibinfo {author} {\bibfnamefont {A.}~\bibnamefont {Gu}}, \bibinfo {author} {\bibfnamefont {T.}~\bibnamefont {Manovitz}}, \bibinfo {author} {\bibfnamefont {M.}~\bibnamefont {Xu}}, \bibinfo {author} {\bibfnamefont {M.}~\bibnamefont {Kalinowski}}, \bibinfo {author} {\bibfnamefont {S.}~\bibnamefont {Majidy}}, \bibinfo {author} {\bibfnamefont {C.}~\bibnamefont {Kokail}}, \bibinfo {author} {\bibfnamefont {N.}~\bibnamefont {Maskara}}, \bibinfo {author} {\bibfnamefont {E.~C.}\ \bibnamefont {Trapp}}, \bibinfo {author} {\bibfnamefont {L.~M.}\ \bibnamefont {Stewart}}, \bibinfo {author} {\bibfnamefont
  {S.}~\bibnamefont {Hollerith}}, \bibinfo {author} {\bibfnamefont {H.}~\bibnamefont {Zhou}}, \bibinfo {author} {\bibfnamefont {M.~J.}\ \bibnamefont {Gullans}}, \bibinfo {author} {\bibfnamefont {S.~F.}\ \bibnamefont {Yelin}}, \bibinfo {author} {\bibfnamefont {M.}~\bibnamefont {Greiner}}, \bibinfo {author} {\bibfnamefont {V.}~\bibnamefont {Vuletić}}, \bibinfo {author} {\bibfnamefont {M.}~\bibnamefont {Cain}},\ and\ \bibinfo {author} {\bibfnamefont {M.~D.}\ \bibnamefont {Lukin}},\ }\bibfield  {title} {\bibinfo {title} {A fault-tolerant neutral-atom architecture for universal quantum computation},\ }\href {https://doi.org/10.1038/s41586-025-09848-5} {\bibfield  {journal} {\bibinfo  {journal} {Nature}\ }\textbf {\bibinfo {volume} {649}},\ \bibinfo {pages} {39} (\bibinfo {year} {2026})}\BibitemShut {NoStop}%
\bibitem [{\citenamefont {Pause}\ \emph {et~al.}(2023)\citenamefont {Pause}, \citenamefont {Preuschoff}, \citenamefont {Sch{\"a}ffner}, \citenamefont {Schlosser},\ and\ \citenamefont {Birkl}}]{Pause2023}%
  \BibitemOpen
  \bibfield  {author} {\bibinfo {author} {\bibfnamefont {L.}~\bibnamefont {Pause}}, \bibinfo {author} {\bibfnamefont {T.}~\bibnamefont {Preuschoff}}, \bibinfo {author} {\bibfnamefont {D.}~\bibnamefont {Sch{\"a}ffner}}, \bibinfo {author} {\bibfnamefont {M.}~\bibnamefont {Schlosser}},\ and\ \bibinfo {author} {\bibfnamefont {G.}~\bibnamefont {Birkl}},\ }\bibfield  {title} {\bibinfo {title} {Reservoir-based deterministic loading of single-atom tweezer arrays},\ }\href {https://doi.org/10.1103/PhysRevResearch.5.L032009} {\bibfield  {journal} {\bibinfo  {journal} {Phys. Rev. Research}\ }\textbf {\bibinfo {volume} {5}},\ \bibinfo {pages} {L032009} (\bibinfo {year} {2023})}\BibitemShut {NoStop}%
\bibitem [{\citenamefont {Norcia}\ \emph {et~al.}(2024)\citenamefont {Norcia}, \citenamefont {Kim}, \citenamefont {Cairncross}, \citenamefont {Stone}, \citenamefont {Ryou}, \citenamefont {Jaffe}, \citenamefont {Brown}, \citenamefont {Barnes}, \citenamefont {Battaglino}, \citenamefont {Bohdanowicz}, \citenamefont {Brown}, \citenamefont {Cassella}, \citenamefont {Chen}, \citenamefont {Coxe}, \citenamefont {Crow}, \citenamefont {Epstein}, \citenamefont {Griger}, \citenamefont {Halperin}, \citenamefont {Hummel}, \citenamefont {Jones}, \citenamefont {Kindem}, \citenamefont {King}, \citenamefont {Kotru}, \citenamefont {Lauigan}, \citenamefont {Li}, \citenamefont {Lu}, \citenamefont {Megidish}, \citenamefont {Marjanovic}, \citenamefont {McDonald}, \citenamefont {Mittiga}, \citenamefont {Muniz}, \citenamefont {Narayanaswami}, \citenamefont {Nishiguchi}, \citenamefont {Paule}, \citenamefont {Pawlak}, \citenamefont {Peng}, \citenamefont {Pudenz}, \citenamefont {Perez}, \citenamefont {Smull}, \citenamefont {Stack},
  \citenamefont {Urbanek}, \citenamefont {van~de Veerdonk}, \citenamefont {Vendeiro}, \citenamefont {Wadleigh}, \citenamefont {Wilkason}, \citenamefont {Wu}, \citenamefont {Xie}, \citenamefont {Zalys-Geller}, \citenamefont {Zhang},\ and\ \citenamefont {Bloom}}]{Norcia2024}%
  \BibitemOpen
  \bibfield  {author} {\bibinfo {author} {\bibfnamefont {M.~A.}\ \bibnamefont {Norcia}}, \bibinfo {author} {\bibfnamefont {H.}~\bibnamefont {Kim}}, \bibinfo {author} {\bibfnamefont {W.~B.}\ \bibnamefont {Cairncross}}, \bibinfo {author} {\bibfnamefont {M.}~\bibnamefont {Stone}}, \bibinfo {author} {\bibfnamefont {A.}~\bibnamefont {Ryou}}, \bibinfo {author} {\bibfnamefont {M.}~\bibnamefont {Jaffe}}, \bibinfo {author} {\bibfnamefont {M.~O.}\ \bibnamefont {Brown}}, \bibinfo {author} {\bibfnamefont {K.}~\bibnamefont {Barnes}}, \bibinfo {author} {\bibfnamefont {P.}~\bibnamefont {Battaglino}}, \bibinfo {author} {\bibfnamefont {T.~C.}\ \bibnamefont {Bohdanowicz}}, \bibinfo {author} {\bibfnamefont {A.}~\bibnamefont {Brown}}, \bibinfo {author} {\bibfnamefont {K.}~\bibnamefont {Cassella}}, \bibinfo {author} {\bibfnamefont {C.~A.}\ \bibnamefont {Chen}}, \bibinfo {author} {\bibfnamefont {R.}~\bibnamefont {Coxe}}, \bibinfo {author} {\bibfnamefont {D.}~\bibnamefont {Crow}}, \bibinfo {author} {\bibfnamefont {J.}~\bibnamefont
  {Epstein}}, \bibinfo {author} {\bibfnamefont {C.}~\bibnamefont {Griger}}, \bibinfo {author} {\bibfnamefont {E.}~\bibnamefont {Halperin}}, \bibinfo {author} {\bibfnamefont {F.}~\bibnamefont {Hummel}}, \bibinfo {author} {\bibfnamefont {A.~M.~W.}\ \bibnamefont {Jones}}, \bibinfo {author} {\bibfnamefont {J.~M.}\ \bibnamefont {Kindem}}, \bibinfo {author} {\bibfnamefont {J.}~\bibnamefont {King}}, \bibinfo {author} {\bibfnamefont {K.}~\bibnamefont {Kotru}}, \bibinfo {author} {\bibfnamefont {J.}~\bibnamefont {Lauigan}}, \bibinfo {author} {\bibfnamefont {M.}~\bibnamefont {Li}}, \bibinfo {author} {\bibfnamefont {M.}~\bibnamefont {Lu}}, \bibinfo {author} {\bibfnamefont {E.}~\bibnamefont {Megidish}}, \bibinfo {author} {\bibfnamefont {J.}~\bibnamefont {Marjanovic}}, \bibinfo {author} {\bibfnamefont {M.}~\bibnamefont {McDonald}}, \bibinfo {author} {\bibfnamefont {T.}~\bibnamefont {Mittiga}}, \bibinfo {author} {\bibfnamefont {J.~A.}\ \bibnamefont {Muniz}}, \bibinfo {author} {\bibfnamefont {S.}~\bibnamefont
  {Narayanaswami}}, \bibinfo {author} {\bibfnamefont {C.}~\bibnamefont {Nishiguchi}}, \bibinfo {author} {\bibfnamefont {T.}~\bibnamefont {Paule}}, \bibinfo {author} {\bibfnamefont {K.~A.}\ \bibnamefont {Pawlak}}, \bibinfo {author} {\bibfnamefont {L.~S.}\ \bibnamefont {Peng}}, \bibinfo {author} {\bibfnamefont {K.~L.}\ \bibnamefont {Pudenz}}, \bibinfo {author} {\bibfnamefont {D.~R.}\ \bibnamefont {Perez}}, \bibinfo {author} {\bibfnamefont {A.}~\bibnamefont {Smull}}, \bibinfo {author} {\bibfnamefont {D.}~\bibnamefont {Stack}}, \bibinfo {author} {\bibfnamefont {M.}~\bibnamefont {Urbanek}}, \bibinfo {author} {\bibfnamefont {R.~J.~M.}\ \bibnamefont {van~de Veerdonk}}, \bibinfo {author} {\bibfnamefont {Z.}~\bibnamefont {Vendeiro}}, \bibinfo {author} {\bibfnamefont {L.}~\bibnamefont {Wadleigh}}, \bibinfo {author} {\bibfnamefont {T.}~\bibnamefont {Wilkason}}, \bibinfo {author} {\bibfnamefont {T.~Y.}\ \bibnamefont {Wu}}, \bibinfo {author} {\bibfnamefont {X.}~\bibnamefont {Xie}}, \bibinfo {author} {\bibfnamefont
  {E.}~\bibnamefont {Zalys-Geller}}, \bibinfo {author} {\bibfnamefont {X.}~\bibnamefont {Zhang}},\ and\ \bibinfo {author} {\bibfnamefont {B.~J.}\ \bibnamefont {Bloom}},\ }\bibfield  {title} {\bibinfo {title} {Iterative assembly of $^{171}${Y}b atom arrays with cavity-enhanced optical lattices},\ }\href@noop {} {\bibfield  {journal} {\bibinfo  {journal} {PRX Quantum}\ }\textbf {\bibinfo {volume} {5}},\ \bibinfo {pages} {030316} (\bibinfo {year} {2024})}\BibitemShut {NoStop}%
\bibitem [{\citenamefont {Gyger}\ \emph {et~al.}(2024)\citenamefont {Gyger}, \citenamefont {Ammenwerth}, \citenamefont {Tao}, \citenamefont {Timme}, \citenamefont {Snigirev}, \citenamefont {Bloch},\ and\ \citenamefont {Zeiher}}]{Gyger2024}%
  \BibitemOpen
  \bibfield  {author} {\bibinfo {author} {\bibfnamefont {F.}~\bibnamefont {Gyger}}, \bibinfo {author} {\bibfnamefont {M.}~\bibnamefont {Ammenwerth}}, \bibinfo {author} {\bibfnamefont {R.}~\bibnamefont {Tao}}, \bibinfo {author} {\bibfnamefont {H.}~\bibnamefont {Timme}}, \bibinfo {author} {\bibfnamefont {S.}~\bibnamefont {Snigirev}}, \bibinfo {author} {\bibfnamefont {I.}~\bibnamefont {Bloch}},\ and\ \bibinfo {author} {\bibfnamefont {J.}~\bibnamefont {Zeiher}},\ }\bibfield  {title} {\bibinfo {title} {Continuous operation of large-scale atom arrays in optical lattices},\ }\href@noop {} {\bibfield  {journal} {\bibinfo  {journal} {Phys. Rev. Res.}\ }\textbf {\bibinfo {volume} {6}},\ \bibinfo {pages} {033104} (\bibinfo {year} {2024})}\BibitemShut {NoStop}%
\bibitem [{\citenamefont {Chiu}\ \emph {et~al.}(2025)\citenamefont {Chiu}, \citenamefont {Trapp}, \citenamefont {Guo}, \citenamefont {Abobeih}, \citenamefont {Stewart}, \citenamefont {Hollerith}, \citenamefont {Stroganov}, \citenamefont {Kalinowski}, \citenamefont {Geim}, \citenamefont {Evered}, \citenamefont {Li}, \citenamefont {Lyu}, \citenamefont {Peters}, \citenamefont {Bluvstein}, \citenamefont {Wang}, \citenamefont {Greiner}, \citenamefont {Vuleti{\'c}},\ and\ \citenamefont {Lukin}}]{Chiu2025}%
  \BibitemOpen
  \bibfield  {author} {\bibinfo {author} {\bibfnamefont {N.-C.}\ \bibnamefont {Chiu}}, \bibinfo {author} {\bibfnamefont {E.~C.}\ \bibnamefont {Trapp}}, \bibinfo {author} {\bibfnamefont {J.}~\bibnamefont {Guo}}, \bibinfo {author} {\bibfnamefont {M.~H.}\ \bibnamefont {Abobeih}}, \bibinfo {author} {\bibfnamefont {L.~M.}\ \bibnamefont {Stewart}}, \bibinfo {author} {\bibfnamefont {S.}~\bibnamefont {Hollerith}}, \bibinfo {author} {\bibfnamefont {P.~L.}\ \bibnamefont {Stroganov}}, \bibinfo {author} {\bibfnamefont {M.}~\bibnamefont {Kalinowski}}, \bibinfo {author} {\bibfnamefont {A.~A.}\ \bibnamefont {Geim}}, \bibinfo {author} {\bibfnamefont {S.~J.}\ \bibnamefont {Evered}}, \bibinfo {author} {\bibfnamefont {S.~H.}\ \bibnamefont {Li}}, \bibinfo {author} {\bibfnamefont {X.}~\bibnamefont {Lyu}}, \bibinfo {author} {\bibfnamefont {L.~M.}\ \bibnamefont {Peters}}, \bibinfo {author} {\bibfnamefont {D.}~\bibnamefont {Bluvstein}}, \bibinfo {author} {\bibfnamefont {T.~T.}\ \bibnamefont {Wang}}, \bibinfo {author} {\bibfnamefont
  {M.}~\bibnamefont {Greiner}}, \bibinfo {author} {\bibfnamefont {V.}~\bibnamefont {Vuleti{\'c}}},\ and\ \bibinfo {author} {\bibfnamefont {M.~D.}\ \bibnamefont {Lukin}},\ }\bibfield  {title} {\bibinfo {title} {Continuous operation of a coherent 3,000-qubit system},\ }\href {https://doi.org/10.1038/s41586-025-09596-6} {\bibfield  {journal} {\bibinfo  {journal} {Nature}\ }\textbf {\bibinfo {volume} {646}},\ \bibinfo {pages} {1075} (\bibinfo {year} {2025})}\BibitemShut {NoStop}%
\bibitem [{\citenamefont {Young}\ \emph {et~al.}(2022)\citenamefont {Young}, \citenamefont {Safari}, \citenamefont {Huft}, \citenamefont {Zhang}, \citenamefont {Oh}, \citenamefont {Chinnarasu},\ and\ \citenamefont {Saffman}}]{Young2022}%
  \BibitemOpen
  \bibfield  {author} {\bibinfo {author} {\bibfnamefont {C.~B.}\ \bibnamefont {Young}}, \bibinfo {author} {\bibfnamefont {A.}~\bibnamefont {Safari}}, \bibinfo {author} {\bibfnamefont {P.}~\bibnamefont {Huft}}, \bibinfo {author} {\bibfnamefont {J.}~\bibnamefont {Zhang}}, \bibinfo {author} {\bibfnamefont {E.}~\bibnamefont {Oh}}, \bibinfo {author} {\bibfnamefont {R.}~\bibnamefont {Chinnarasu}},\ and\ \bibinfo {author} {\bibfnamefont {M.}~\bibnamefont {Saffman}},\ }\bibfield  {title} {\bibinfo {title} {An architecture for quantum networking of neutral atom processors},\ }\href@noop {} {\bibfield  {journal} {\bibinfo  {journal} {Appl.Phys. B}\ }\textbf {\bibinfo {volume} {128}},\ \bibinfo {pages} {151} (\bibinfo {year} {2022})}\BibitemShut {NoStop}%
\bibitem [{\citenamefont {Singh}\ \emph {et~al.}(2023)\citenamefont {Singh}, \citenamefont {Bradley}, \citenamefont {Anand}, \citenamefont {Ramesh}, \citenamefont {White},\ and\ \citenamefont {Bernien}}]{Singh2023}%
  \BibitemOpen
  \bibfield  {author} {\bibinfo {author} {\bibfnamefont {K.}~\bibnamefont {Singh}}, \bibinfo {author} {\bibfnamefont {C.~E.}\ \bibnamefont {Bradley}}, \bibinfo {author} {\bibfnamefont {S.}~\bibnamefont {Anand}}, \bibinfo {author} {\bibfnamefont {V.}~\bibnamefont {Ramesh}}, \bibinfo {author} {\bibfnamefont {R.}~\bibnamefont {White}},\ and\ \bibinfo {author} {\bibfnamefont {H.}~\bibnamefont {Bernien}},\ }\bibfield  {title} {\bibinfo {title} {Mid-circuit correction of correlated phase errors using an array of spectator qubits},\ }\href@noop {} {\bibfield  {journal} {\bibinfo  {journal} {Science}\ }\textbf {\bibinfo {volume} {380}},\ \bibinfo {pages} {1265} (\bibinfo {year} {2023})}\BibitemShut {NoStop}%
\bibitem [{\citenamefont {Miles}\ \emph {et~al.}(2026)\citenamefont {Miles}, \citenamefont {Lichtman}, \citenamefont {Scott}, \citenamefont {Scott}, \citenamefont {Norrell}, \citenamefont {Bedalov}, \citenamefont {Belknap}, \citenamefont {Cole}, \citenamefont {Eubanks}, \citenamefont {Gillette}, \citenamefont {Gokhale}, \citenamefont {Goldwin}, \citenamefont {Iliev}, \citenamefont {Jones}, \citenamefont {Kuper}, \citenamefont {Mason}, \citenamefont {Mitchell}, \citenamefont {Murphree}, \citenamefont {Neff-Mallon}, \citenamefont {Noel}, \citenamefont {Radnaev}, \citenamefont {Vinogradov},\ and\ \citenamefont {Saffman}}]{Miles2026}%
  \BibitemOpen
  \bibfield  {author} {\bibinfo {author} {\bibfnamefont {J.}~\bibnamefont {Miles}}, \bibinfo {author} {\bibfnamefont {M.~T.}\ \bibnamefont {Lichtman}}, \bibinfo {author} {\bibfnamefont {A.~M.}\ \bibnamefont {Scott}}, \bibinfo {author} {\bibfnamefont {J.}~\bibnamefont {Scott}}, \bibinfo {author} {\bibfnamefont {S.~A.}\ \bibnamefont {Norrell}}, \bibinfo {author} {\bibfnamefont {M.~J.}\ \bibnamefont {Bedalov}}, \bibinfo {author} {\bibfnamefont {D.~A.}\ \bibnamefont {Belknap}}, \bibinfo {author} {\bibfnamefont {D.~C.}\ \bibnamefont {Cole}}, \bibinfo {author} {\bibfnamefont {S.~Y.}\ \bibnamefont {Eubanks}}, \bibinfo {author} {\bibfnamefont {M.}~\bibnamefont {Gillette}}, \bibinfo {author} {\bibfnamefont {P.}~\bibnamefont {Gokhale}}, \bibinfo {author} {\bibfnamefont {J.}~\bibnamefont {Goldwin}}, \bibinfo {author} {\bibfnamefont {M.}~\bibnamefont {Iliev}}, \bibinfo {author} {\bibfnamefont {R.~A.}\ \bibnamefont {Jones}}, \bibinfo {author} {\bibfnamefont {K.~W.}\ \bibnamefont {Kuper}}, \bibinfo {author} {\bibfnamefont
  {D.}~\bibnamefont {Mason}}, \bibinfo {author} {\bibfnamefont {P.~T.}\ \bibnamefont {Mitchell}}, \bibinfo {author} {\bibfnamefont {J.~D.}\ \bibnamefont {Murphree}}, \bibinfo {author} {\bibfnamefont {N.~A.}\ \bibnamefont {Neff-Mallon}}, \bibinfo {author} {\bibfnamefont {T.~W.}\ \bibnamefont {Noel}}, \bibinfo {author} {\bibfnamefont {A.~G.}\ \bibnamefont {Radnaev}}, \bibinfo {author} {\bibfnamefont {I.~V.}\ \bibnamefont {Vinogradov}},\ and\ \bibinfo {author} {\bibfnamefont {M.}~\bibnamefont {Saffman}},\ }\bibfield  {title} {\bibinfo {title} {Qubit syndrome measurements with a high fidelity {R}b-{C}s {R}ydberg gate},\ }\href {https://arxiv.org/abs/2603.13492} {\bibfield  {journal} {\bibinfo  {journal} {arXiv:2603.13492}\ } (\bibinfo {year} {2026})}\BibitemShut {NoStop}%
\bibitem [{\citenamefont {Li}\ \emph {et~al.}(2026)\citenamefont {Li}, \citenamefont {Dilley}, \citenamefont {Gonzales}, \citenamefont {Hahn}, \citenamefont {White}, \citenamefont {Arnon}, \citenamefont {Bernien}, \citenamefont {Saleem},\ and\ \citenamefont {Jiang}}]{BLi2026}%
  \BibitemOpen
  \bibfield  {author} {\bibinfo {author} {\bibfnamefont {B.}~\bibnamefont {Li}}, \bibinfo {author} {\bibfnamefont {D.}~\bibnamefont {Dilley}}, \bibinfo {author} {\bibfnamefont {A.}~\bibnamefont {Gonzales}}, \bibinfo {author} {\bibfnamefont {T.~A.}\ \bibnamefont {Hahn}}, \bibinfo {author} {\bibfnamefont {R.}~\bibnamefont {White}}, \bibinfo {author} {\bibfnamefont {R.}~\bibnamefont {Arnon}}, \bibinfo {author} {\bibfnamefont {H.}~\bibnamefont {Bernien}}, \bibinfo {author} {\bibfnamefont {Z.}~\bibnamefont {Saleem}},\ and\ \bibinfo {author} {\bibfnamefont {L.}~\bibnamefont {Jiang}},\ }\bibfield  {title} {\bibinfo {title} {Efficient entanglement purification circuit design for dual-species atom arrays},\ }\href {https://doi.org/10.1103/6j15-7mty} {\bibfield  {journal} {\bibinfo  {journal} {PRX Quantum}\ } (\bibinfo {year} {2026})}\BibitemShut {NoStop}%
\bibitem [{\citenamefont {Chou}\ \emph {et~al.}(2005)\citenamefont {Chou}, \citenamefont {de~Riedmatten}, \citenamefont {Felinto}, \citenamefont {Polyakov}, \citenamefont {van Enk},\ and\ \citenamefont {Kimble}}]{Chou2005}%
  \BibitemOpen
  \bibfield  {author} {\bibinfo {author} {\bibfnamefont {C.~W.}\ \bibnamefont {Chou}}, \bibinfo {author} {\bibfnamefont {H.}~\bibnamefont {de~Riedmatten}}, \bibinfo {author} {\bibfnamefont {D.}~\bibnamefont {Felinto}}, \bibinfo {author} {\bibfnamefont {S.~V.}\ \bibnamefont {Polyakov}}, \bibinfo {author} {\bibfnamefont {S.~J.}\ \bibnamefont {van Enk}},\ and\ \bibinfo {author} {\bibfnamefont {H.~J.}\ \bibnamefont {Kimble}},\ }\bibfield  {title} {\bibinfo {title} {Measurement-induced entanglement for excitation stored in remote atomic ensembles},\ }\href {https://doi.org/10.1038/nature04353} {\bibfield  {journal} {\bibinfo  {journal} {Nature}\ }\textbf {\bibinfo {volume} {438}},\ \bibinfo {pages} {828} (\bibinfo {year} {2005})}\BibitemShut {NoStop}%
\bibitem [{Hwa()}]{Hwang2026SM}%
  \BibitemOpen
  \href@noop {} {}\bibinfo {note} {See Supplemental Material at [URL will be inserted by publisher] which includes references \cite{Safari2026,Volz2006,Wilk2007,vanLeent2020,Zhang2022,Hartung2024,LLi2025,Thompson2013b,Young2022,garthoff2021efficient,Kuhr2005,Saffman2011}}\BibitemShut {NoStop}%
\bibitem [{\citenamefont {Brown}\ and\ \citenamefont {Twiss}(1956)}]{HanburyBrown1956}%
  \BibitemOpen
  \bibfield  {author} {\bibinfo {author} {\bibfnamefont {R.~H.}\ \bibnamefont {Brown}}\ and\ \bibinfo {author} {\bibfnamefont {R.~Q.}\ \bibnamefont {Twiss}},\ }\bibfield  {title} {\bibinfo {title} {Correlation between photons in two coherent beams of light},\ }\href@noop {} {\bibfield  {journal} {\bibinfo  {journal} {Nature}\ }\textbf {\bibinfo {volume} {177}},\ \bibinfo {pages} {27} (\bibinfo {year} {1956})}\BibitemShut {NoStop}%
\bibitem [{\citenamefont {Bennett}\ \emph {et~al.}(1996)\citenamefont {Bennett}, \citenamefont {DiVincenzo}, \citenamefont {Smolin},\ and\ \citenamefont {Wootters}}]{Bennett1996}%
  \BibitemOpen
  \bibfield  {author} {\bibinfo {author} {\bibfnamefont {C.~H.}\ \bibnamefont {Bennett}}, \bibinfo {author} {\bibfnamefont {D.~P.}\ \bibnamefont {DiVincenzo}}, \bibinfo {author} {\bibfnamefont {J.~A.}\ \bibnamefont {Smolin}},\ and\ \bibinfo {author} {\bibfnamefont {W.~K.}\ \bibnamefont {Wootters}},\ }\bibfield  {title} {\bibinfo {title} {Mixed-state entanglement and quantum error correction},\ }\href@noop {} {\bibfield  {journal} {\bibinfo  {journal} {Phys. Rev. A}\ }\textbf {\bibinfo {volume} {54}},\ \bibinfo {pages} {3824} (\bibinfo {year} {1996})}\BibitemShut {NoStop}%
\bibitem [{\citenamefont {Kucera}\ \emph {et~al.}(2024)\citenamefont {Kucera}, \citenamefont {Haen}, \citenamefont {Arensk{\"o}tter}, \citenamefont {Bauer}, \citenamefont {Meiers}, \citenamefont {Sch{\"a}fer}, \citenamefont {Boland}, \citenamefont {Yahyapour}, \citenamefont {Lessing}, \citenamefont {Holzwarth}, \citenamefont {Becher},\ and\ \citenamefont {Eschner}}]{Kucera2024}%
  \BibitemOpen
  \bibfield  {author} {\bibinfo {author} {\bibfnamefont {S.}~\bibnamefont {Kucera}}, \bibinfo {author} {\bibfnamefont {C.}~\bibnamefont {Haen}}, \bibinfo {author} {\bibfnamefont {E.}~\bibnamefont {Arensk{\"o}tter}}, \bibinfo {author} {\bibfnamefont {T.}~\bibnamefont {Bauer}}, \bibinfo {author} {\bibfnamefont {J.}~\bibnamefont {Meiers}}, \bibinfo {author} {\bibfnamefont {M.}~\bibnamefont {Sch{\"a}fer}}, \bibinfo {author} {\bibfnamefont {R.}~\bibnamefont {Boland}}, \bibinfo {author} {\bibfnamefont {M.}~\bibnamefont {Yahyapour}}, \bibinfo {author} {\bibfnamefont {M.}~\bibnamefont {Lessing}}, \bibinfo {author} {\bibfnamefont {R.}~\bibnamefont {Holzwarth}}, \bibinfo {author} {\bibfnamefont {C.}~\bibnamefont {Becher}},\ and\ \bibinfo {author} {\bibfnamefont {J.}~\bibnamefont {Eschner}},\ }\bibfield  {title} {\bibinfo {title} {Demonstration of quantum network protocols over a 14-km urban fiber link},\ }\href {https://doi.org/10.1038/s41534-024-00886-x} {\bibfield  {journal} {\bibinfo  {journal} {npj Qu. Inf.}\
  }\textbf {\bibinfo {volume} {10}},\ \bibinfo {pages} {88} (\bibinfo {year} {2024})}\BibitemShut {NoStop}%
\bibitem [{\citenamefont {Blinov}\ \emph {et~al.}(2004)\citenamefont {Blinov}, \citenamefont {Moehring}, \citenamefont {Duan},\ and\ \citenamefont {Monroe}}]{Blinov04}%
  \BibitemOpen
  \bibfield  {author} {\bibinfo {author} {\bibfnamefont {B.~B.}\ \bibnamefont {Blinov}}, \bibinfo {author} {\bibfnamefont {D.~L.}\ \bibnamefont {Moehring}}, \bibinfo {author} {\bibfnamefont {L.~M.}\ \bibnamefont {Duan}},\ and\ \bibinfo {author} {\bibfnamefont {C.}~\bibnamefont {Monroe}},\ }\bibfield  {title} {\bibinfo {title} {Observation of entanglement between a single trapped atom and a single photon},\ }\href@noop {} {\bibfield  {journal} {\bibinfo  {journal} {Nature}\ }\textbf {\bibinfo {volume} {428}},\ \bibinfo {pages} {153} (\bibinfo {year} {2004})}\BibitemShut {NoStop}%
\bibitem [{\citenamefont {Anand}\ \emph {et~al.}(2024)\citenamefont {Anand}, \citenamefont {Bradley}, \citenamefont {White}, \citenamefont {Ramesh}, \citenamefont {Singh},\ and\ \citenamefont {Bernien}}]{Anand2024}%
  \BibitemOpen
  \bibfield  {author} {\bibinfo {author} {\bibfnamefont {S.}~\bibnamefont {Anand}}, \bibinfo {author} {\bibfnamefont {C.~E.}\ \bibnamefont {Bradley}}, \bibinfo {author} {\bibfnamefont {R.}~\bibnamefont {White}}, \bibinfo {author} {\bibfnamefont {V.}~\bibnamefont {Ramesh}}, \bibinfo {author} {\bibfnamefont {K.}~\bibnamefont {Singh}},\ and\ \bibinfo {author} {\bibfnamefont {H.}~\bibnamefont {Bernien}},\ }\bibfield  {title} {\bibinfo {title} {A dual-species {R}ydberg array},\ }\href@noop {} {\bibfield  {journal} {\bibinfo  {journal} {Nat. Phys.}\ }\textbf {\bibinfo {volume} {20}},\ \bibinfo {pages} {1744} (\bibinfo {year} {2024})}\BibitemShut {NoStop}%
\bibitem [{\citenamefont {Wilk}\ \emph {et~al.}(2007{\natexlab{b}})\citenamefont {Wilk}, \citenamefont {Webster}, \citenamefont {Specht}, \citenamefont {Rempe},\ and\ \citenamefont {Kuhn}}]{Wilk2007}%
  \BibitemOpen
  \bibfield  {author} {\bibinfo {author} {\bibfnamefont {T.}~\bibnamefont {Wilk}}, \bibinfo {author} {\bibfnamefont {S.~C.}\ \bibnamefont {Webster}}, \bibinfo {author} {\bibfnamefont {H.~P.}\ \bibnamefont {Specht}}, \bibinfo {author} {\bibfnamefont {G.}~\bibnamefont {Rempe}},\ and\ \bibinfo {author} {\bibfnamefont {A.}~\bibnamefont {Kuhn}},\ }\bibfield  {title} {\bibinfo {title} {Polarization-controlled single photons},\ }\href@noop {} {\bibfield  {journal} {\bibinfo  {journal} {Phys. Rev. Lett.}\ }\textbf {\bibinfo {volume} {98}},\ \bibinfo {pages} {063601} (\bibinfo {year} {2007}{\natexlab{b}})}\BibitemShut {NoStop}%
\bibitem [{\citenamefont {Thompson}\ \emph {et~al.}(2013)\citenamefont {Thompson}, \citenamefont {Tiecke}, \citenamefont {Zibrov}, \citenamefont {Vuleti\ifmmode~\acute{c}\else \'{c}\fi{}},\ and\ \citenamefont {Lukin}}]{Thompson2013b}%
  \BibitemOpen
  \bibfield  {author} {\bibinfo {author} {\bibfnamefont {J.~D.}\ \bibnamefont {Thompson}}, \bibinfo {author} {\bibfnamefont {T.~G.}\ \bibnamefont {Tiecke}}, \bibinfo {author} {\bibfnamefont {A.~S.}\ \bibnamefont {Zibrov}}, \bibinfo {author} {\bibfnamefont {V.}~\bibnamefont {Vuleti\ifmmode~\acute{c}\else \'{c}\fi{}}},\ and\ \bibinfo {author} {\bibfnamefont {M.~D.}\ \bibnamefont {Lukin}},\ }\bibfield  {title} {\bibinfo {title} {Coherence and {R}aman sideband cooling of a single atom in an optical tweezer},\ }\href@noop {} {\bibfield  {journal} {\bibinfo  {journal} {Phys. Rev. Lett.}\ }\textbf {\bibinfo {volume} {110}},\ \bibinfo {pages} {133001} (\bibinfo {year} {2013})}\BibitemShut {NoStop}%
\bibitem [{\citenamefont {Garthoff}(2021)}]{garthoff2021efficient}%
  \BibitemOpen
  \bibfield  {author} {\bibinfo {author} {\bibfnamefont {R.}~\bibnamefont {Garthoff}},\ }\href {https://books.google.com/books?id=uw3LzgEACAAJ} {\emph {\bibinfo {title} {Efficient Single Photon Collection for Single Atom Quantum Nodes}}}\ (\bibinfo  {publisher} {M{\"u}nchen, Ludwig-Maximilians-Universit{\"a}t},\ \bibinfo {year} {2021})\BibitemShut {NoStop}%
\bibitem [{\citenamefont {Kuhr}\ \emph {et~al.}(2005)\citenamefont {Kuhr}, \citenamefont {Alt}, \citenamefont {Schrader}, \citenamefont {Dotsenko}, \citenamefont {Miroshnychenko}, \citenamefont {Rauschenbeutel},\ and\ \citenamefont {Meschede}}]{Kuhr2005}%
  \BibitemOpen
  \bibfield  {author} {\bibinfo {author} {\bibfnamefont {S.}~\bibnamefont {Kuhr}}, \bibinfo {author} {\bibfnamefont {W.}~\bibnamefont {Alt}}, \bibinfo {author} {\bibfnamefont {D.}~\bibnamefont {Schrader}}, \bibinfo {author} {\bibfnamefont {I.}~\bibnamefont {Dotsenko}}, \bibinfo {author} {\bibfnamefont {Y.}~\bibnamefont {Miroshnychenko}}, \bibinfo {author} {\bibfnamefont {A.}~\bibnamefont {Rauschenbeutel}},\ and\ \bibinfo {author} {\bibfnamefont {D.}~\bibnamefont {Meschede}},\ }\bibfield  {title} {\bibinfo {title} {Analysis of dephasing mechanisms in a standing-wave dipole trap},\ }\href@noop {} {\bibfield  {journal} {\bibinfo  {journal} {Phys. Rev. A}\ }\textbf {\bibinfo {volume} {72}},\ \bibinfo {pages} {023406} (\bibinfo {year} {2005})}\BibitemShut {NoStop}%
\bibitem [{\citenamefont {Saffman}\ \emph {et~al.}(2011)\citenamefont {Saffman}, \citenamefont {Zhang}, \citenamefont {Gill}, \citenamefont {Isenhower},\ and\ \citenamefont {Walker}}]{Saffman2011}%
  \BibitemOpen
  \bibfield  {author} {\bibinfo {author} {\bibfnamefont {M.}~\bibnamefont {Saffman}}, \bibinfo {author} {\bibfnamefont {X.~L.}\ \bibnamefont {Zhang}}, \bibinfo {author} {\bibfnamefont {A.~T.}\ \bibnamefont {Gill}}, \bibinfo {author} {\bibfnamefont {L.}~\bibnamefont {Isenhower}},\ and\ \bibinfo {author} {\bibfnamefont {T.~G.}\ \bibnamefont {Walker}},\ }\bibfield  {title} {\bibinfo {title} {Rydberg state mediated quantum gates and entanglement of pairs of neutral atoms},\ }\href@noop {} {\bibfield  {journal} {\bibinfo  {journal} {J. Phys.: Conf. Ser.}\ }\textbf {\bibinfo {volume} {264}},\ \bibinfo {pages} {012023} (\bibinfo {year} {2011})}\BibitemShut {NoStop}%
\end{thebibliography}%
